\documentclass[amsmath,amssymb,aps,prl,reprint,floatfix]{revtex4-1}

\usepackage{graphicx}
\usepackage{dcolumn}
\usepackage{bm}
\usepackage[none]{hyphenat}
\usepackage[mathlines]{lineno}
\usepackage{float}
\usepackage [english]{babel}
\usepackage [autostyle, english = american]{csquotes}
\usepackage[utf8]{inputenc}
\MakeOuterQuote{"}

\newcommand{\ket}[1]{\left|#1\right\rangle}
\newcommand{\down}{\downarrow}
\newcommand{\up}{\uparrow}
\newcommand{\broket}[3]{\left\langle #1|#2|#3\right\rangle}
\newcommand{\braket}[2]{\left\langle #1|#2\right\rangle}
\newcommand{\ord}[1]{\mathcal{O}\left(#1\right)}
\newcommand{\ketbra}[2]{\left|#1\right\rangle\left\langle#2\right|}
\newcommand{\rom}[1]{\uppercase\expandafter{\romannumeral #1\relax}}

\begin{document}

\title{Measuring magnetic fields with magnetic field insensitive transitions}

\author{Yotam Shapira}
\email{yotam.shapira@weizmann.ac.il}
\author{Yehonatan Dallal}
\author{Roee Ozeri}
\author{Ady Stern}
\affiliation{%
Department of Physics, Weizmann Institute of Science, Rehovot 7610001, Israel
}%

\date{\today}

\begin{abstract}
Magnetometry is an important tool prevalent in many applications such as fundamental research, material characterization and biological imaging. Atomic magnetometry conventionally makes use of two quantum states, the energy difference of which depends linearly on the magnetic field due to the Zeeman effect. The magnetic field is evaluated from repeated measurements of the accumulated dynamic phase between the two Zeeman states in a superposition. Here we propose a magnetometry method that employs a superposition of clock states with energies that do not depend, to first-order, on the magnetic field magnitude. Our method makes use of the geometrical dependence of the clock-states wavefunctions on the magnetic field orientation. We propose detailed schemes for measuring both static and time-varying magnetic fields, and analyze the sensitivity of these methods. We show that, similarly to Zeeman-based methods, the smallest measurable signal scales inversely with the system coherence-time, which for clock transitions is typically significantly longer than for magnetically sensitive transitions. Finally, we experimentally demonstrate our method on an ensemble of optically trapped $^{87}\text{Rb}$ atoms.
\end{abstract}

\maketitle
\section{I. Introduction}
Magnetometry is widely used in many diverse fields, including material characterization \cite{Romalis2011}, geomagnetic surveys \cite{Nabighian2005,Mathe2005}, tests of fundamental physics \cite{Berglund1995,Altarev2009,Lee2018}, biological imaging \cite{Bison2003,Belfi2007} and more. Contemporary high-sensitivity magnetometers, demonstrating sensitivities below one $\text{ fT}/\sqrt{\text{Hz}}$, typically make use of superconducting quantum interference devices (SQUID) \cite{Kirtley1999,Robbes2006}, or atomic systems. The most sensitive magnetometer demonstrated to-date is the spin-exchange relaxation-free (SERF) atomic magnetometer \cite{Allred2002,Sheng2013}.

Conventionally, atomic magnetometry is performed by tracking an accumulated dynamical phase of a magnetic-field-dependent transition of choice, which evolves due to Larmor precession, and comparing it to a stable local oscillator; e.g the Zeeman ground state manifold of $^{87}\text{Rb}$ atoms compared to a driving RF field. Such a system was originally proposed by Dehmelt \cite{Dehmelt1957} and demonstrated for the first time by Bell and Bloom \cite{Bell1957,Bloom1962}.

Similarly, atomic clocks operate by locking a local oscillator, an optical or RF source, to a transition frequency between two quantum states. Since stability is a crucial property of any clock, the atomic states are typically chosen such that the transition is as insensitive as possible to ambient magnetic fields \cite{Essen1955,Yu1992}. The transition between such \textit{clock states} is typically first-order insensitive to magnetic fields. A good example is the $\ket{6^2 S_{1/2},F=4,m_F=0}\leftrightarrow\ket{6^2 S_{1/2},F=3,m_F=0}$ transition in the $^{133}$Cs atom on which the SI second is defined \cite{Terrien1968}.

Here we propose a geometric atomic magnetometry method which employs exactly such clock states, despite the insensitivity of their transition to the ambient magnetic field. We show that even in the absence of a dynamically evolving phase, geometric orientation alone may be used to evaluate magnetic fields. Crucially, the method's sensitivity is ultimately limited by the clock state's coherence time, which is significantly longer as compared with that of the more conventional Zeeman-split states.

Our article is structured as follows. Section \rom{2} highlights the essential physical picture and our main results. Section \rom{3} introduces a two spin-$\frac{1}{2}$ toy-model which captures most features of our method. Section \rom{4} applies our method to a more practical hyperfine atomic system and derives a Hamiltonian which corresponds to the toy model in Sec. \rom{3}. In section \rom{5} we derive in detail DC magnetometry in the atomic system and propose a Ramsey-like magnetometry scheme. Section \rom{6} generalizes our method to AC magentometry. Section \rom{7} provides a detailed analysis of our method's sensitivity and compares it to conventional methods. Four appendices provide further details of calculations. 

\section{II. Physical picture and main results}
Here we measure magnetic fields using a superposition of clock states. While it is true that, to first order, the transition frequency between these states is unaffected by the magnetic field magnitude, their wavefunctions are changed by a rotation of the magnetic field. Changes in the magnetic field direction, due to a magnetic field component which is perpendicular to the externally applied quantization field, result in a change in the coupling between the clock states using a third driving field. This change in coupling can be translated to a change in state-populations, which is linear in the perpendicular field magnitude. Our magnetometry method thus measures the perpendicular field magnitude.

The method is captured, in essence, by a two spin-$\frac{1}{2}$ toy-model system, subject to a Zeeman-like Hamiltonian. The two states with $S_z=0$, namely the singlet $\ket{S}=\frac{\ket{\up\down}-\ket{\down\up}}{\sqrt{2}}$ and triplet $m=0$, $\ket{T}=\frac{\ket{\up\down}+\ket{\down\up}}{\sqrt{2}}$, are degenerate and therefore their transition energy is magnetic field independent. Here $\ket{\up}$ and $\ket{\down}$ refer to the single-spin eigenstates, pointing along the magnetic field direction, $\hat{b}$. The other two triplet states, which are fully polarized, are gaped in energy. We assume the energy separation to be large enough such that these states may be ignored. Then, the $\ket{S}$ and $\ket{T}$ states can be described as the two opposing poles of a Bloch sphere, which we describe by an "iso-spin" $\bf\tau$. Since both states do not have a real orientation, the sphere does not represent any real direction in space. 

\begin{figure}[t]\includegraphics[width=\columnwidth]{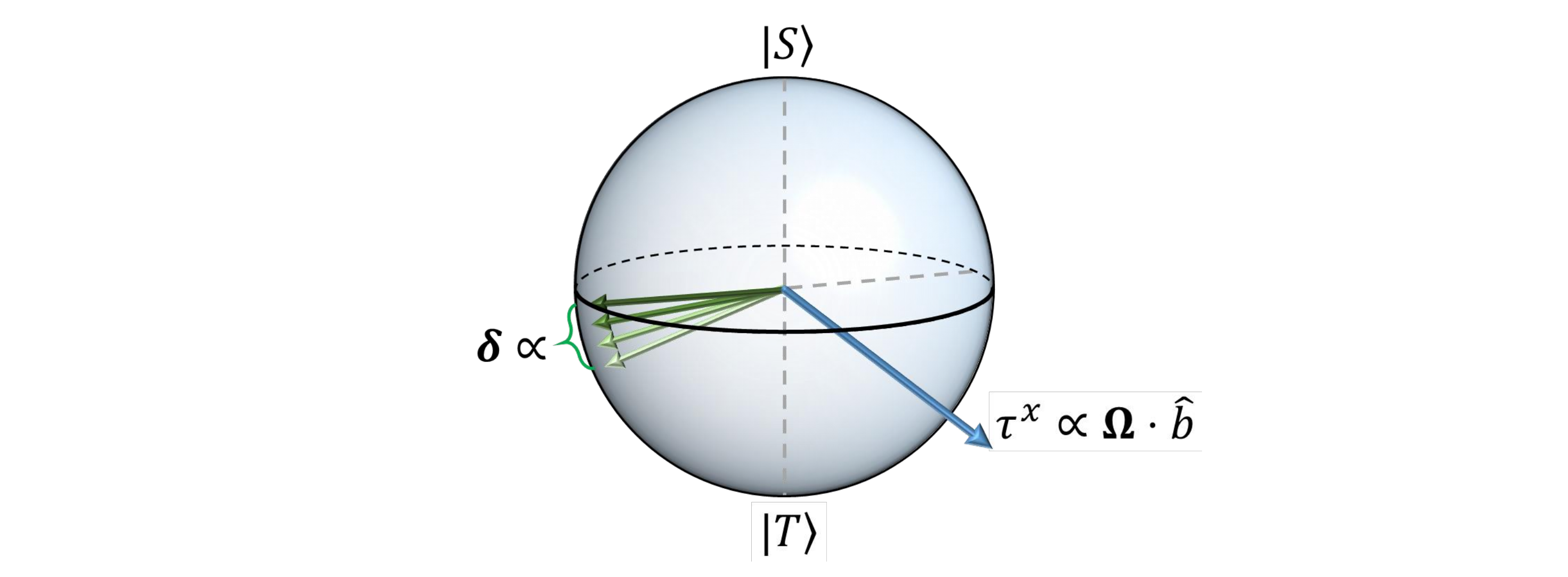}
\caption{\textbf{Bloch sphere representing all superpositions of $\ket{S}$ and $\ket{T}$ .} The singlet and $m=0$ triplet states, $\ket{S}$ and $\ket{T}$, degenerate under an externally applied magnetic field $\hat{b}$, are represented by vectors pointing to the sphere poles. $\tau^x$ rotations (blue arrow) are implemented by applying a magnetic field, $\boldsymbol{\Omega}$, exclusively to the first spin. The rotation is then proportional to the projection of $\boldsymbol{\Omega}$ on $\hat{b}$. A state which is initially in an equal superposition of $\ket{S}$ and $\ket{T}$ (dark green) will rotate toward one of the poles at an angle which is proportional to $\boldsymbol{\Omega}\cdot\hat{b}$ (lighter green).}\label{figBloch}\end{figure}

A rotation of the state-vector in this Bloch sphere may be implemented via a local interaction with only one of the spins, e.g. $\boldsymbol{\Omega}\cdot\boldsymbol{\sigma}_1$, where $\boldsymbol{\sigma}_1=\left(\sigma^x_1,\sigma^y_1,\sigma^z_1\right)$ and $\boldsymbol{\Omega}$ is an additional driving field. Remarkably, it is only the component parallel to the quantization field,  $\boldsymbol{\Omega}\cdot\hat{b}$, that couples $\ket{S}$ and $\ket{T}$, and acts like a $\tau_x$ operator within the clock states subspace. The other components of $\bf\Omega$ flip $\sigma_1^z$ and therefore couple $\ket{S}$ and $\ket{T}$ to the fully polarized states, which are energetically largely separated. Figure \ref{figBloch} shows the clock subspace Bloch sphere and the $\tau^x$ rotation operator acting on an equal superposition state.

Magnetometry in the clock subspace is possible by initializing the system in the state $\frac{\ket{S}-i\ket{T}}{\sqrt{2}}$ and directing $\boldsymbol{\Omega}$ such that it is perpendicular to the direction of the externally applied quantization field, $\hat{b}$. If the quantization field is the only one present, then the $\boldsymbol{\Omega}$-drive is decoupled from the clock subspace, and the population in the $\ket{S}$ state remains $\frac{1}{2}$. However any additional magnetic field ("signal"), $\boldsymbol{\delta}$, which is parallel to $\boldsymbol{\Omega}$, and therefore perpendicular to $\hat{b}$, will rotate the superposition in the clock subspace leading to change in the population of $\ket{S}$ which is linear in the signal field amplitude (green arrows in Fig. \ref{figBloch}).

The realization of this toy model requires state preparation and measurement in the entangled Bell-basis which is often challenging. Furthermore, it limits the dynamics on the Bloch sphere by allowing simple application of $\tau_x$ operations only. To alleviate these two difficulties we extend our method to a hyperfine atomic system. Here the two clock states are no longer degenerate. Thus they are coupled with an AC driving field, which renders them degenerate, in leading order of the rotating wave approximation. The polarization of this field becomes another degree of freedom in the problem. Specifically, in the rotating frame the phase-difference between the two polarization "arms" allows for the application of $\bf\tau$ rotation operators around any direction in the $x-y$ plane of the Bloch sphere.

To measure magnetic fields using hyperfine clock superpositions we initialize the system in one of the clock states represented by a pole on the Bloch sphere (e.g., $\tau_z=+1$) and apply a microwave pulse to rotate around the $\tau_x$ direction to an equal superposition of clock states pointing in the $\tau_y$-axis direction. Then we apply a second phase-shifted microwave pulse such that, in the absence of a "signal" field, it applies a $\tau_y$ rotation to the superposition, and therefore does not change the state in the clock subspace up to a global phase. Any signal field which is perpendicular to the quantization axis and lies in the plane defined by the polarization ellipse of the microwave field will cause the second microwave pulse to rotates the $\tau$ spin away from an equal superposition, leading again to a change in clock state population which is linear in the signal field magnitude.

We investigate both DC and AC magnetometry schemes and provide several experimental verifications of the validity of our proposal, performed on an ensemble of optically trapped $^{87}\text{Rb}$ atoms.

Finally, we analyze  the sensitivity of our method. We find that the smallest measurable signal scales inversely with the system coherence time. Such a scaling may be expected, since the coherence time is the only time scale that ultimately limits the measurement. This scale also sets the fundamental limit to Zeeman-based magnetometry. However, clock state coherence times are typically significantly longer compared to coherence times of Zeeman-split states \cite{Langer2005,Kleine2011}, leading to potentially increased sensitivities. 

\section{III. Two spin-$\frac{1}{2}$ system}
We investigate the Hamiltonian of two non-interacting spin-$\frac{1}{2}$ particles, where one of the spins is interacting with an additional, time-dependent, magnetic field,
\begin{equation}
H=\mu\boldsymbol{B}\left(t\right)\cdot\left(\boldsymbol{\sigma}_{1}+\boldsymbol{\sigma}_{2}\right)+\hbar\boldsymbol{\Omega}\left(t\right)\cdot\boldsymbol{\sigma}_{1}. \label{eqnHam}
\end{equation}
Here $\boldsymbol{\sigma}_{i}=\left(\sigma_{i}^{x},\sigma_{i}^{y},\sigma_{i}^{z}\right)$ are the Pauli operators acting on the $i$'th spin, $\boldsymbol{B}=B\hat{b}$ is the ambient magnetic field, $\mu$ is a magneton-like coupling constant and $\boldsymbol{\Omega}$ is an additional driving field acting exclusively on the first spin. We assume $\boldsymbol{\Omega}$ lies in the $\hat{x}-\hat{z}$ plane, at an angle $\chi$ to the $\hat{z}$ axis.

For $\boldsymbol{\Omega}=0$ and $\boldsymbol{B}=B_{z}\hat{z}$ the Hamiltonian in Eq. \eqref{eqnHam} is diagonalized by the eigenstates $\ket{\up\up}$ and $\ket{\down\down}$ with energies $\pm \mu B$, and in addition by $\ket{S}\equiv\frac{\ket{\up\down}-\ket{\down\up}}{\sqrt{2}}$ and $\ket{T}\equiv\frac{\ket{\up\down}+\ket{\down\up}}{\sqrt{2}}$, denoting the singlet and triplet $m=0$ states, with vanishing energy, and therefore a magnetic field insensitive transition energy. Nevertheless we will show that the $\ket{S}\leftrightarrow\ket{T}$ transition can be used for precise magnetometry.

We turn on $\boldsymbol{\Omega}$ adiabatically in a constant direction, while Maintaining $\boldsymbol{B}$ in the $\hat{z}$ direction, such that we keep $\hbar^2\frac{\partial\Omega}{\partial t}\ll\mu^2 B^2$ and $\hbar\Omega\ll\mu B$. This ensures that the driving field can only induce transitions in the clock subspace. This clock subspace and driving Hamiltonian can therefore be described on a Bloch sphere with $\ket{S}$ and $\ket{T}$ on the north and south poles respectively.

Since $\broket{S}{\sigma_1^j}{S}=\broket{T}{\sigma_1^j}{T}=0$ and $\broket{S}{\sigma_1^j}{T}=\delta_{j,z}$ only the $\Omega_z$ component can introduce transitions between $\ket{S}$ and $\ket{T}$. The matrix element is real so these transitions are Pauli-$x$ like rotations. To avoid ambiguity we shall denote this driving as a $\tau^x$ rotation on the Bloch sphere. Indeed, acting with this drive such that $\int\Omega\left(t\right)dt=\frac{\pi}{2}$ we rotate $\ket{S}$ to $\frac{\ket{S}-i\ket{T}}{\sqrt{2}}$ pointing in the $-\hat{y}$ direction. When $\boldsymbol{B}$ points to an arbitrary direction the effective Hamiltonian in the clock subspace due to the drive generalizes to $\hbar\Omega_z\tau^x\rightarrow\hbar\left(\boldsymbol{\Omega}\cdot\hat{b}\right)\tau^x$. To complete the Bloch sphere picture, we note that the Casimir operator, $\sum_i\sigma_1^i\sigma_2^i$, acts as a $\tau^z$ in the clock subspace and $\tau^y\equiv-i\tau^z\tau^x$. 

\begin{figure}\includegraphics[width=\columnwidth]{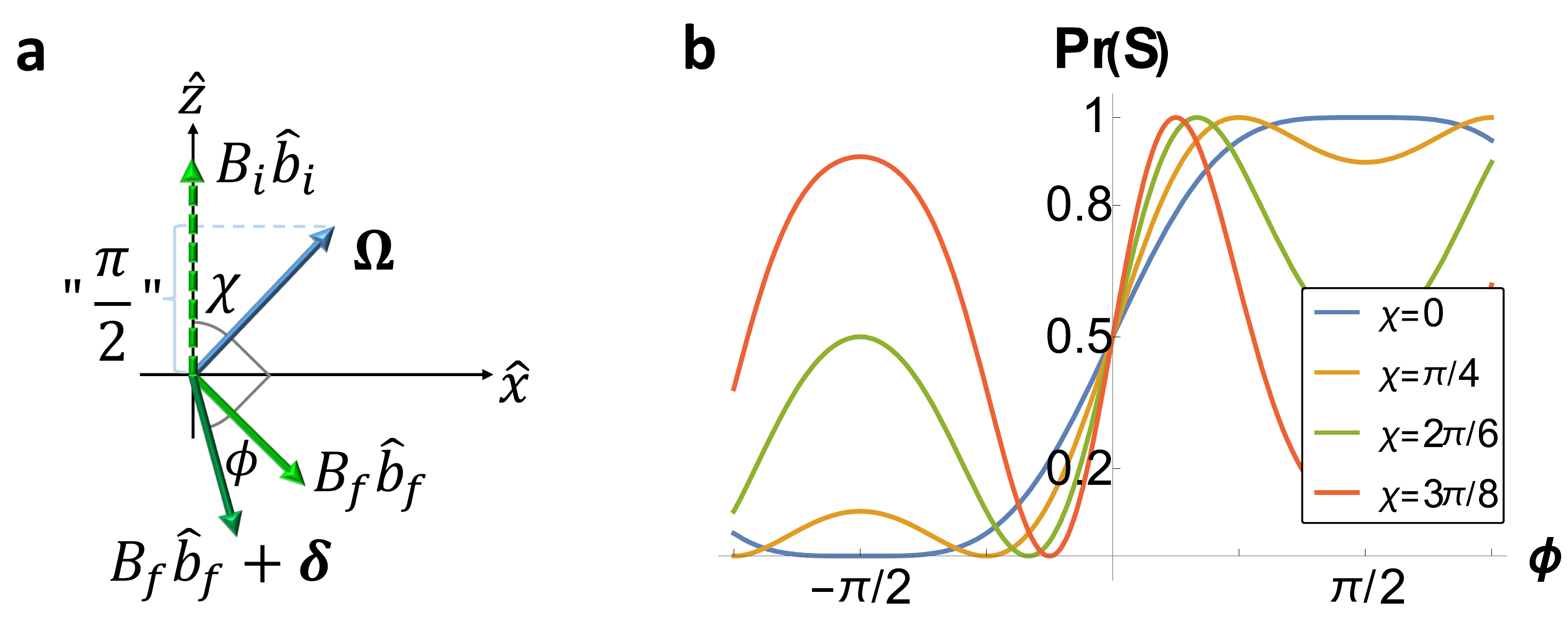}
\caption{\textbf{Magnetometry scheme using the clock states of a two spin-$\frac{1}{2}$ system. (a)} Scheme layout. The magnetic field is initialized on the $\hat{b}_i=\hat{z}$ direction (dashed green) with magnitude $B_i$. A pulse is applied using the driving field, $\boldsymbol{\Omega}$ (blue), which is oriented at angle $\chi$ with respect to $\hat{z}$, and implements a $\frac{\pi}{2}$-pulse. Then $\boldsymbol{B}$ is decreased to $B_f$, such that $B_i\gg B_f\gg\delta$, and rotated such that $\boldsymbol{\Omega}\perp\hat{b}_f$ (green), a component of $\boldsymbol{\delta}$ which is parallel to $\boldsymbol{\Omega}$ will rotate the total field by an angle $\phi$ with respect to $\hat{b}_f$ (dark green). A second $\frac{\pi}{2}$-pulse will measure this rotation. \textbf{(b)} Probability of finding the system in the $\ket{S}$ state after the second pulse as a function of $\phi$ for different drive orientations, $\chi$, according to Eq. \eqref{eqnPrSpinFull}. All configurations show a sensitivity set by a linear slope around $\phi=0$ which extends over a finite range of $\phi$. As $\chi$ approaches $\frac{\pi}{2}$ the measurement sensitivity increases and its dynamic range decreases. At the $\chi=\pi/2$ point $\Omega$ and $B$ are orthogonal and therefore the necessary $\Omega$ for a $\pi/2$ pulse diverges.}\label{FigTwoSpins}\end{figure}

To detect a small magnetic field $\boldsymbol{\delta}=\left(\delta_{x},\delta_{y},\delta_{z}\right)$ we use the drive twice at different orientations of $\boldsymbol{B}$. As shown below, any deviation from an equal superposition of $\ket{S}$ and $\ket{T}$ will be due to $\boldsymbol{\delta}$.

Specifically, we start with $\boldsymbol{B}=B_i\hat{z}$ such that $B_i\gg\delta$, and initialize the system to $\ket{S}$. We turn on $\boldsymbol{\Omega}\left(t\right)$ adiabatically such that a pulse is applied in the $\hat{\Omega}$ direction, for convenience we define the pulse area as $\Omega T$. Next we decrease the magnetic field magnitude adiabatically to $B_f$ such that $B_i\gg B_f\gg\delta$, and rotate the direction of the field from $\hat{z}$ to $\hat{b}_f$. The total magnetic field is now directed towards
\begin{equation}
    \frac{B_f\hat{b}_f+\boldsymbol{\delta}}{\left|B_f\hat{b}_f+\boldsymbol{\delta}\right|}=\hat{b}_f+\frac{\boldsymbol{\delta}}{B_f}-\frac{\boldsymbol{\delta}\cdot\hat{b}_f}{B_f}\hat{b}_f+\mathcal{O}\left(\frac{\delta^{2}}{B_f^{2}}\right).\label{eqnBexp}  
\end{equation}

We perform an additional adiabatic pulse, which is identical to the first one. Since both pulses constitute a $\tau^{x}$ rotation then after these two pulses the probability of measuring the system in the $\ket{S}$ state is 
\begin{equation}
\text{Pr}\left(S\right)\approx\cos^{2}\left(\left(\hat{z}+\hat{b}_f+\frac{\boldsymbol{\delta}}{B_f}-\frac{\boldsymbol{\delta}\cdot\hat{b}_f}{B_f}\hat{b}_f\right)\cdot\frac{\boldsymbol{\Omega}T}{2}\right). \label{eqnPrS} 
\end{equation}

Making the choice $\hat{b}_f\cdot\hat{\Omega}=0$, i.e the pulse is orthogonal to the final magnetic field direction, then only the $\boldsymbol{\delta}\cdot\hat{\Omega}$ term affects the population probability. We expand the cosine term and obtain,
\begin{equation}
\text{Pr}\left(S\right)=\cos^{2}\left(\hat{z}\cdot\frac{\boldsymbol{\Omega}T}{2}\right)-2\sin\left(\hat{z}\cdot\boldsymbol{\Omega}T\right)\frac{\boldsymbol{\delta}}{B_f}\cdot\frac{\boldsymbol{\Omega}T}{2}+\mathcal{O}\left(\frac{\delta^{2}}{B_f^{2}}\right). \label{eqnPrSExp}    
\end{equation}

To maximize the sensitivity we maximize the sine term in Eq. \eqref{eqnPrSExp} by setting $\Omega_{z}=\frac{\pi}{2T}$. That is, $\Omega_{z}$ is such that the first pulse is always a $\frac{\pi}{2}$-pulse. For simplicity we assume that $\boldsymbol{\delta}$ lies in the $\hat{x}-\hat{z}$ plane, then we may scan different components of $\boldsymbol{\delta}$ by varying the angle $\chi$, between $\boldsymbol{\Omega}$ and the $\hat{z}$ direction. We note that as $\chi$ approaches $\frac{\pi}{2}$ the required amplitude of $\boldsymbol{\Omega}$ diverges. The probability to find the system in $\ket{S}$ is, in leading order, $\text{Pr}\left(S\right)=\frac{1}{2}-\frac{\boldsymbol{\delta}\cdot\boldsymbol{\Omega}}{B_f}$. Any deviation from $\text{Pr}\left(S\right)=\frac{1}{2}$ may now be attributed to a component $\boldsymbol{\delta}$ along the direction of $\boldsymbol{\Omega}$. Since both $\boldsymbol{\Omega}$ and $\boldsymbol{B}$ are controlled, this constitutes a measurement of $\delta$. Figure \ref{FigTwoSpins}a shows the geometrical layout of this scheme.

Assuming $\boldsymbol{\delta}$ acts to rotate the total magnetic field in Eq. \eqref{eqnBexp} by an angle $\phi$, then the exact probability to find the system in the $\ket{S}$ state is given by
\begin{equation}
    \text{Pr}\left(S\right)=\sin^{2}\left(\frac{\pi}{4}\left(1+\frac{\sin\left(\phi\right)}{\cos\left(\chi\right)}\right)\right).\label{eqnPrSpinFull}
\end{equation}
Figure \ref{FigTwoSpins}b shows the probability to find the system in the $\ket{S}$ state due to a rotation of the total magnetic field by an angle $\phi$ as a consequence of $\boldsymbol{\delta}$, for different choices of orientations of $\boldsymbol{\Omega}$. Clearly as $\chi$ approaches $\frac{\pi}{2}$ the sensitivity increases, seen by the increasing slope around $\phi=0$, while the measurement range decreases, seen by the approach of the extremum points towards $\phi=0$. 

Alternatively we may measure the magnetic field $\delta_x$ by only reducing the magnetic field amplitude from $B_i$ to $B_f$, while keeping it at the $\hat{z}$ direction. Following a similar two-pulse scheme, with $\Omega T=\frac{\pi}{4}$, i.e $\frac{\pi}{4}$-pulses, the probability to remain in the singlet state is,
\begin{equation}
    \text{Pr}\left(S\right)=\cos^{2}\left(\left(1+\frac{\cos\left(\chi-\phi\right)}{\cos\left(\chi\right)}\right)\frac{\pi}{8}\right)\approx\frac{1}{2}-\frac{1}{2}\frac{\Omega_{x}\delta_{x}}{B}T,\label{eqnPrSpinStat}
\end{equation}
which has a similar structure as Eq. \eqref{eqnPrSExp} and \eqref{eqnPrSpinFull}, yet with half the sensitivity.

We pause to highlight the essence of our method. The $\ket{S}$ and $\ket{T}$ states are degenerate, their transition energy is not affected by the magnetic field and therefore there is no dynamical phase difference that can be measured when interfering them. Naively one would expect the $\ket{S}\leftrightarrow\ket{T}$ transition to be useless for magnetometry. However $\ket{T}$ is affected by the magnetic field, it is oriented by it and defined in terms of spin-states along its direction. Formally this effect is due to the scalar product between $\boldsymbol{B}$ and the Pauli matrices. Here we have utilized this dependence in order to measure changes in the magnetic field orientation. Our method consists of two identical pulses after which, in the absence of any additional magnetic field, $\delta$, the system is in an equal superposition of the $\ket{S}$ and $\ket{T}$ states. Any deviation from such a superposition is then mapped to a rotation angle of the total magnetic field and to the component of $\delta$ orthogonal to $\hat{b}_f$. The magnetic field is effectively sampled at the second $\frac{\pi}{2}$-pulse instance, thus the measurement is insensitive to the magnetic field trajectory between the two pulses.

So far our discussion and derivations have been in the "magnetic" frame, i.e the frame in which changes in the magnetic field direction do not affect the $\ket{S}$ and $\ket{T}$ states. It is also possible to repeat the discussion above in the static "lab" frame, in which the system states is given in the $\sigma_1^z+\sigma_2^z$ eigenstates basis. Clearly this approach is equivalent and reproduces the same probability as in Eq. \eqref{eqnPrSpinFull}, however it is less intuitive (see appendix A).

\section{IV. Hyperfine clock states}
Taking this method to a more realistic system, we consider a clock transition between two hyperfine states in the ground state of an alkali atom. For concreteness we focus on the eight states of the $F=1$ and $F=2$ hyperfine manifolds of the $5S_{1/2}$ ground level of $^{87}\text{Rb}$. At zero magnetic field, the $\ket{F=1,m_F=0}$ and $\ket{F=2,m_F=0}$ are clock states, with their transition energy being insensitive to the magnetic field, to leading order. 

We experimentally demonstrated our methods on a cloud of ultra-cold $^{87}\text{Rb}$ atoms. The atoms were collected from a magneto-optical trap and then evaporatively cooled to $\approx30\ \mu$K in a $\text{CO}_2$ laser quasielectrostatic trap. We drove the transition between the $F=1$ and $F=2$ hyperfine manifolds using a microwave antenna, tuned to the 6.8 GHz resonance frequency of this transition. The atoms were prepared in the $\ket{1,0}$ state using optical-pumping pulses on the $\ket{F=1}\rightarrow\ket{F=2^\prime}$ $D_2$ transition combined with microwave pulses. The population in the $\ket{2,0}$ state was measured using absorption imaging of the $\ket{2,0}$ state normalized by absorption imaging of all the atoms in both the $F=1$ and $F=2$ manifolds. Further information regarding the setup may be found in \cite{Dallal2014,Dallal2015}.

To analyze such a system we consider the lab-frame Hamiltonian in the intermediate magnetic field regime,
\begin{equation}
    \begin{cases}
H=H_{HF}+H_{Z}+V\left(t\right)\\
H_{HF}=\frac{\hbar A_{HF}}{2}\boldsymbol{I}\cdot\boldsymbol{J}\\
H_{Z}=\mu_{N}g_I\boldsymbol{B}\cdot\boldsymbol{I}+\mu_{B}g_J\boldsymbol{B}\cdot\boldsymbol{J}\\
V=\hbar\left(\frac{\boldsymbol{\Omega}}{2}e^{i\omega_{RF}t}+h.c\right)\cdot\left(\mu_{N}g_I\boldsymbol{I}+\mu_{B}g_J\boldsymbol{J}\right)
\end{cases}\label{eqnHamLab},
\end{equation}
where $H_{HF}$ is the hyperfine interaction Hamiltonian, which couples the nucleus spin operators $\boldsymbol{I}$ with the electronic spin operators $\boldsymbol{J}$ such that the hyperfine splitting is $A_{HF}$. The term $H_Z$ is the Zeeman Hamiltonian, describing the coupling of the quantization field $\boldsymbol{B}$ to the nuclear and electronic spins through their respective Bohr magnetons, $\mu_N$ and $\mu_B$, and the Landé g-factors, $g_I$ and $g_J$. The third term, $V$, describes the same Zeeman coupling to an additional time-dependent, RF, magnetic field used to drive transitions between the two clock states. In order to consider an RF field with a general polarization, we assume that $\boldsymbol{\Omega}$ is complex; i.e. that the RF drive can be written as two orthogonal quadratures of the RF field. For simplicity we restrict $\boldsymbol{\Omega}$ such the resulting polarization ellipse lies in a plane containing the quantization field direction.

The hyperfine Hamiltonian in Eq. \eqref{eqnHamLab} can be diagonalized in the $\ket{F,m_F}$ basis. By shifting it appropriately it becomes $H_{HF}=\frac{\hbar A_{HF}}{2}\left(\delta_{F,2}-\delta_{F,1}\right)$. Choosing $m_F$ along the direction of $\boldsymbol{B}$, $H_Z$ can be written as a direct sum of five subspaces marked by their $m_F$ values, $H_{Z}=H_{Z}^{m_{F}=-2}\oplus H_{Z}^{m_{F}=-1}\oplus...\oplus H_{Z}^{m_{F}=2}$. In the clock subspace the Zeeman Hamiltonian is, $H_{Z}^{m_{F}=0}=\frac{\mu B}{2}\tau^x$, with $\mu\equiv g_I\mu_{N}-g_J\mu_{B}$. 
 
When $\omega_{RF}$ is tuned close to the clock transition frequency and far-detuned from all other transitions (compared to $\left|\boldsymbol{\Omega}\right|$), we can assume it does not excite any transitions outside of the clock states subspace. 

The lab-frame Hamiltonian in the clock subspace is therefore composed of the hyperfine splitting, a Zeeman term and the RF drive,
\begin{equation}
H_{\text{clk,lab}}=\frac{\hbar A_{HF}}{2}\tau^{z}+\frac{1}{2}\left(\mu B+\hbar\left(\Omega_{z}e^{i\omega_{RF}t}+h.c\right)\right)\tau^{x}.\label{eqnHamClkLab}
\end{equation}
The $B$-dependent Zeeman term weakly mixes the $\ket{2,0}$ and $\ket{1,0}$ states, resulting in a small energy shift, which is quadratic in $\frac{\mu B}{\hbar A_{HF}}$ and is known as the second-order Zeeman shift. In leading order the $m_F=0$ states are clock states. For $^{87}\text{Rb}$ the ground state hyperfine frequency splitting is approximately $6.8\text{ GHz}$ while the Zeeman splitting in these manifolds is approximately $\pm\Delta m\cdot0.70\text{ MHz/Gauss}$ \cite{Steck2001}, justifying our approximation for a wide range of magnetic field magnitudes. 

For a general magnetic field $\Omega_z=\boldsymbol{\Omega}\cdot\hat{b}$ We change to a rotating frame with respect to $\frac{\hbar\omega_{RF}}{2}\tau^z$, and perform a rotating wave approximation, neglecting terms rotating with rate $\omega_{RF}$ or faster, to obtain the interaction picture Hamiltonian,
\begin{equation}
H_{clk,I}=\frac{\hbar\eta}{2}\tau^{z}+\frac{\hbar}{2}\left(\boldsymbol{\Omega}\cdot\hat{b}\tau^{+}+h.c\right), \label{eqnHamRot}
\end{equation}
where $\eta=A_{HF}-\omega_{RF}$ is the RF drive detuning. This frame is diagonal in the measurement basis, therefore we can freely choose $\ket{2,0}$ as the initial state. Equation \eqref{eqnHamRot} shows that the phase of the Rabi frequency is sensitive to the projection of the RF field on the magnetic field direction.

For a general elliptically polarized RF field, $\boldsymbol{\Omega}=e^{i\theta}\left(\boldsymbol{\Omega}_1+i\boldsymbol{\Omega}_2\right)$, where $\boldsymbol{\Omega}_1$ and $\boldsymbol{\Omega}_2$ are the major and minor orthogonal axes of the polarization ellipse and $\theta$ is the RF phase, the on-resonance ($\eta=0$) Hamiltonian is
\begin{equation} 
\begin{cases}
H_{\text{clk}}\left(\hat{b},\theta\right)=\frac{\hbar}{2}\Omega_\text{eff}\left(\cos\left(\xi\right)\tau^x+\sin\left(\xi\right)\tau^y\right)\\
\Omega_\text{eff}=\sqrt{\left(\boldsymbol{\Omega}_{1}\cdot\hat{b}\right)^{2}+\left(\boldsymbol{\Omega}_{2}\cdot\hat{b}\right)^{2}}\\
\xi=\theta+\arctan\left(\frac{\boldsymbol{\Omega}_{2}\cdot\hat{b}}{\boldsymbol{\Omega}_{1}\cdot\hat{b}}\right)
\end{cases}\label{eqnHamClk},
\end{equation}
where we have arbitrarily defined $\tau^x$ as the rotation operator acting at $\theta=0$ (as observables can only be sensitive to phase differences between different pulses). Equation \eqref{eqnHamClk} gives rise to a Ramsey-like Hamiltonian where $\xi$ play the role of the Ramsey-pulse phase, i.e it is the angle between the $\hat{x}$ axis and the Rabi vector on Bloch sphere equator. 

\begin{figure}\includegraphics[width=\columnwidth]{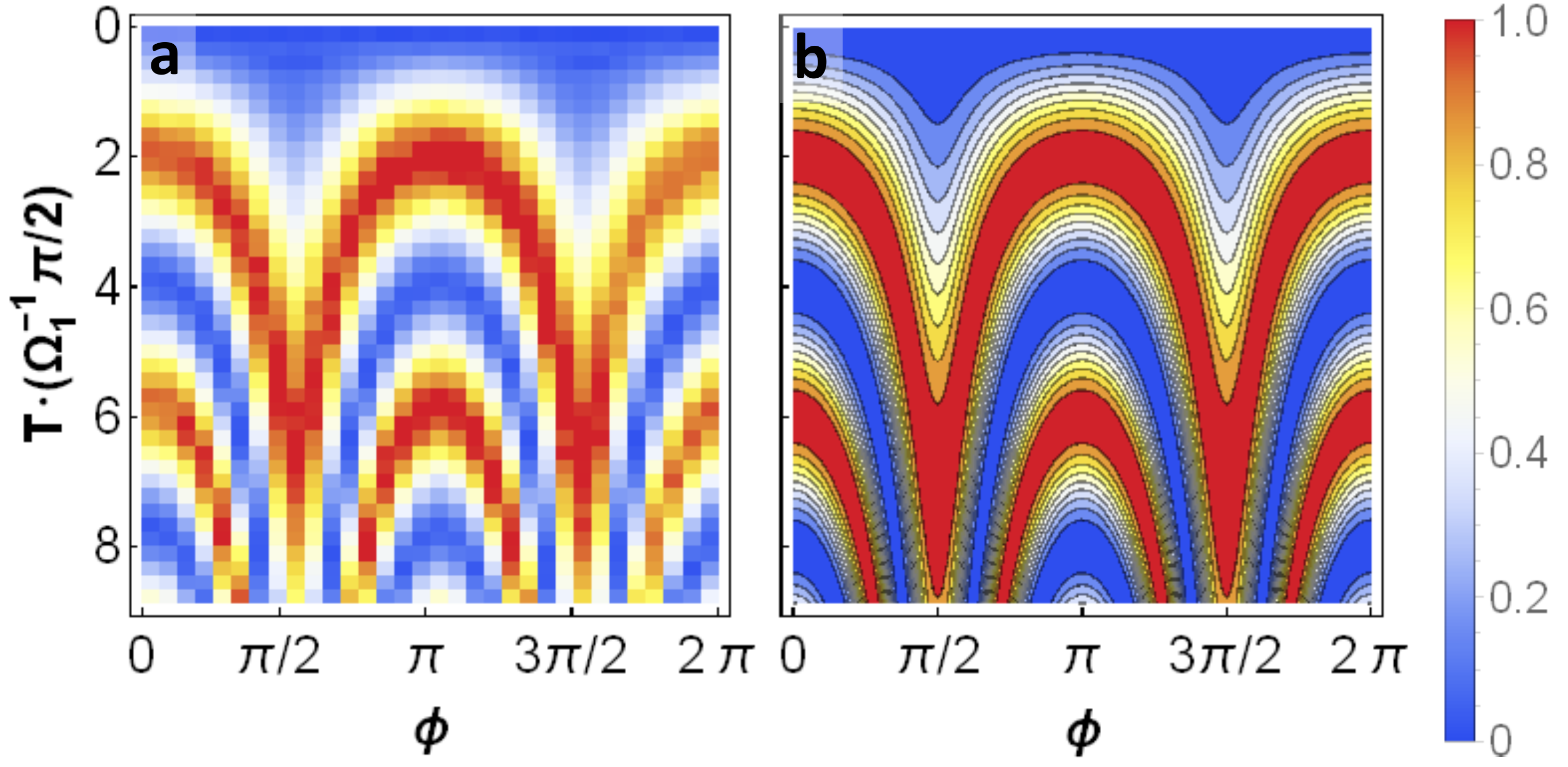}
\caption{\textbf{Rabi oscillations between the $\ket{2,0}$ and $\ket{1,0}$ clock states for varying pulse duration and magnetic field angles. (a)} Data obtained by driving an ensemble of $^{87}\text{Rb}$ atoms with an RF drive, on-resonance with the $\ket{2,0}\leftrightarrow\ket{1,0}$ transition. The angle $\phi=0$ is defined as the angle in which the population of $\ket{2,0}$ maximizes in the shortest time. \textbf{(b)} Theoretical model according to Eq. \eqref{eqnRabi}, with setting the minor-to-major polarization ellipse axis ratio to $\tilde{\Omega}=0.27$ in order to fit the experimental data. Clearly the model fits nicely to the data.}\label{FigRotScan}\end{figure}

Before describing our magnetometry scheme we show how the Hamiltonian in Eq. \eqref{eqnHamClk} generates Rabi nutation of population between the two clock states. Here $\hat{b}$ lies in the polarization ellipse plane at an angle $\phi$ with $\boldsymbol{\Omega_1}$. This further simplifies Eq. \eqref{eqnHamClk} such that $\xi=\theta+\phi$. The probability for the system to remain in $\ket{2,0}$ state, $P_2$, is then given by,
\begin{equation}
    \begin{cases}
    P_2=\sin^{2}\left(\frac{\Omega_{1}T}{2}\beta\right)\\
    \beta\equiv\sqrt{\cos^{2}\left(\phi\right)+\tilde{\Omega}^{2}\sin^{2}\left(\phi\right)}\\
    \tilde{\Omega}=\Omega_2/\Omega_1
    \end{cases}.\label{eqnRabi}
\end{equation}
As expected from a single pulse measurement, Eq. \eqref{eqnRabi} is independent of the RF phase $\theta$. We show below that much like the angle $\chi$, appearing in the two spin-$\frac{1}{2}$ model in Sec. \rom{3}, the dimensionless minor-to-major polarization ellipse axis ratio, $\tilde{\Omega}$, acts as a sensitivity "knob" for the geometrical magnetoemtry method. Figure \ref{FigRotScan}a shows measured Rabi oscillations between the $\ket{2,0}$ and $\ket{1,0}$ clock states, for varying pulse duration, $T$, and magnetic field angle, $\phi$. The angle $\phi=0$ is defined as the magnetic field direction in which the pulse duration that fully excites the system to the $\ket{1,0}$ state is shortest. Our data is in good agreement with the model in Eq. \eqref{eqnRabi} shown in Fig. \ref{FigRotScan}b.

\section{V. DC magnetometry}
Similar to the two spin-$\frac{1}{2}$ magnetometry scheme above, we may initialize the system to $\ket{2,0}$ with $\boldsymbol{B}=B_i\hat{z}$ such that $B_i\gg\delta$ and perform an on-resonance pulse for time $T_1$, then slowly rotate the field to $\hat{b}_f$ while decreasing its magnitude to $B_f$, such that $B_i\gg B_f\gg\delta$, and perform another pulse for time $T_2$. The probability to remain in $\ket{2,0}$ is derived by expanding the magnetic field direction according to Eq. \eqref{eqnBexp}. 

To map the two spin-$\frac{1}{2}$ system in Eq. \eqref{eqnHam} to the atomic system we choose $\boldsymbol{\Omega}_1\cdot\hat{b}=0$ and $\boldsymbol{\Omega}_2=0$ (i.e linear polarization). This recovers exactly the two spin-$\frac{1}{2}$ system results in Eq. \eqref{eqnPrS} and \eqref{eqnPrSpinFull}.

The atomic hyperfine system allows for a more practical way to perform magnetometry, which does not require a rotation of the magnetic field, but rather uses the two polarization axes and $\theta$, the phase of the RF drive. By fixing $\hat{b}$, aligning the polarization ellipse major axis with the magnetic field, $\boldsymbol{\Omega}_1 T_1=\frac{\pi}{2}\hat{b}\Rightarrow\boldsymbol{\Omega}_2\cdot\hat{b}=0$, and setting the second RF pulse phase to $\theta=\frac{\pi}{2}$, the probability to remain in $\ket{2,0}$ becomes
\begin{equation}
    P_2=\frac{1}{2}+\frac{1}{2}\sin\left(\Omega_\text{eff}T_{2}\right)\tilde{\Omega}\frac{\delta}{B_f}\hat{\Omega}_2\cdot\hat{\delta}\label{eqnMagHF}
\end{equation}

Equation \eqref{eqnMagHF} implies that any deviation from $P_2=\frac{1}{2}$ is due to a component of $\boldsymbol{\delta}$ which is parallel to $\boldsymbol{\Omega_2}$, i.e lying on the polarization ellipse and perpendicular to $\hat{b}$. For simplicity and without loss of generality we assume from this point forward that this is the only component of $\boldsymbol{\delta}$, and that $T_2=T_1$. The sensitivity is linear in the minor-to-major ratio $\tilde{\Omega}$, and the measurement range is given by, $\delta_{\text{max}}=\frac{B_{f}}{\tilde{\Omega}}$, as the total rotation due to $\frac{\delta}{B_f}$ has to be small in terms of $\tilde{\Omega}$ for this approximation to hold. 

\begin{figure}[t]\includegraphics[width=\columnwidth]{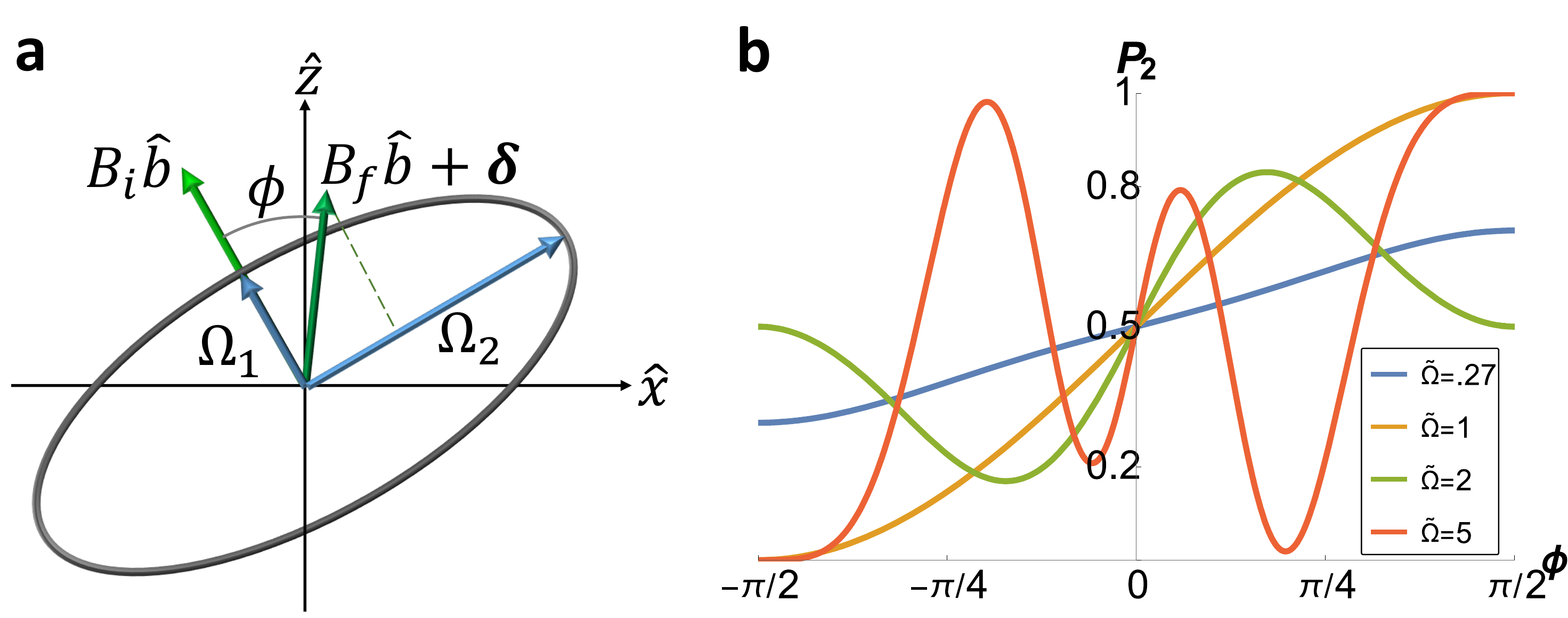}
\caption{\textbf{Magnetometry scheme using the clock states of a hyperfine atomic system. (a)} Scheme layout. The polarization ellipse (solid gray) is aligned such that the major axis $\boldsymbol{\Omega}_1$ is parallel to the magnetic field direction $\hat{b}$ (green). When lowering the field magnitude from $B_i$ to $B_f$, the total magnetic field rotates by an angle $\phi$ (dark green). For the RF phase, $\theta=\frac{\pi}{2}$, the $\ket{2,0}$ population is determined by the projection of the rotated magnetic field on $\Omega_2$ (dashed green). \textbf{(b)} Population in the $\ket{2,0}$ state, $P_2$, as a function of the magnetic field rotation angle $\phi$, according to Eq. \eqref{eqnPr20}, with $\theta=\frac{\pi}{2}$. Clearly as the minor-to-major ratio, $\tilde{\Omega}$, increases the sensitivity increases, seen as an increased slope, while the measurement range, $\delta_{\text{max}}$, decreases due to the decreased separation between the two extrema around $\phi=0$.}\label{FigSens}\end{figure}

Assuming the magnetic field rotation due to $\delta$ is by an angle $\phi$ we may exactly calculate the probability of the system to remain in $\ket{2,0}$. It is given by
\begin{equation}
    P_2=\frac{1}{2}-\frac{\left|\cos\left(\theta\right)\cos\left(\phi\right)-\tilde{\Omega}\sin\left(\theta\right)\sin\left(\phi\right)\right|}{2\beta}\sin\left(\frac{\pi}{2}\beta\right),\label{eqnPr20}
\end{equation}
with $\beta$ defined in Eq. \eqref{eqnRabi}. Figure \ref{FigSens}a shows the geometrical setup of the measurement method. Figure \ref{FigSens}b shows the probability to remain in the $\ket{2,0}$ state due to a rotation of the magnetic field, $\phi$, according to Eq. \eqref{eqnPr20}, with $\theta=\frac{\pi}{2}$. The sensitivity vs. range trade-off is apparent, as a larger minor-to-major axes ratio increases the slope around $\phi=0$, while reducing the distance between the curve maxima and minima.

By considering the clock subspace Bloch sphere the method becomes intuitive. The two axes of the polarization ellipse are orthogonal to one another in quadrature, the first pulse acts as a $\tau^x$ rotation due to the major arm, rotating the state to the $-\hat{y}$ direction. In the absence of any perpendicular magnetic field the $\frac{\pi}{2}$ phase-shifted second pulse acts as a $\tau^y$ rotation and does not affect the state. If however a perpendicular magnetic field component does exist then the state is rotated again by a $\tau^x$ operator, with a coupling strength equal to the magnetic field projection onto the ellipse's minor arm. A too large perpendicular component will over-rotate the state, giving rise to the minima and maxima in Fig. \ref{FigSens} limiting the sensitivity range.  

Determining the magnetic field through a single change in population may suffer from systematic biases caused by experimental imperfections (e.g state preparation errors). To mitigate these effects we make use of a Ramsey fringe measurement analogue. Instead of fixing the second $\frac{\pi}{2}$-pulse phase, $\theta$, we scan it and, according to Eq. \eqref{eqnPr20}, obtain a fringe. We define the fringe phase, $\theta_f$ as the second pulse phase that maximizes the population in the $\ket{2,0}$ state. Analytically it is given by
\begin{equation}
    \theta_f=\pi-\text{sgn}\left(\phi\right)\text{arccos}\left(\cos\left(\phi\right)/\beta\right),\label{eqnThf}
\end{equation}
with $\beta$ defined in Eq. \eqref{eqnRabi}. For $\Omega_2=\Omega_1$, we get $\beta=1$, and therefore the fringe phase is simply $\phi$, and in the limit $\Omega_2\gg\Omega_1$ the fringe converges to a step function around $\phi=0$.

\begin{figure}[t]\includegraphics[width=\columnwidth]{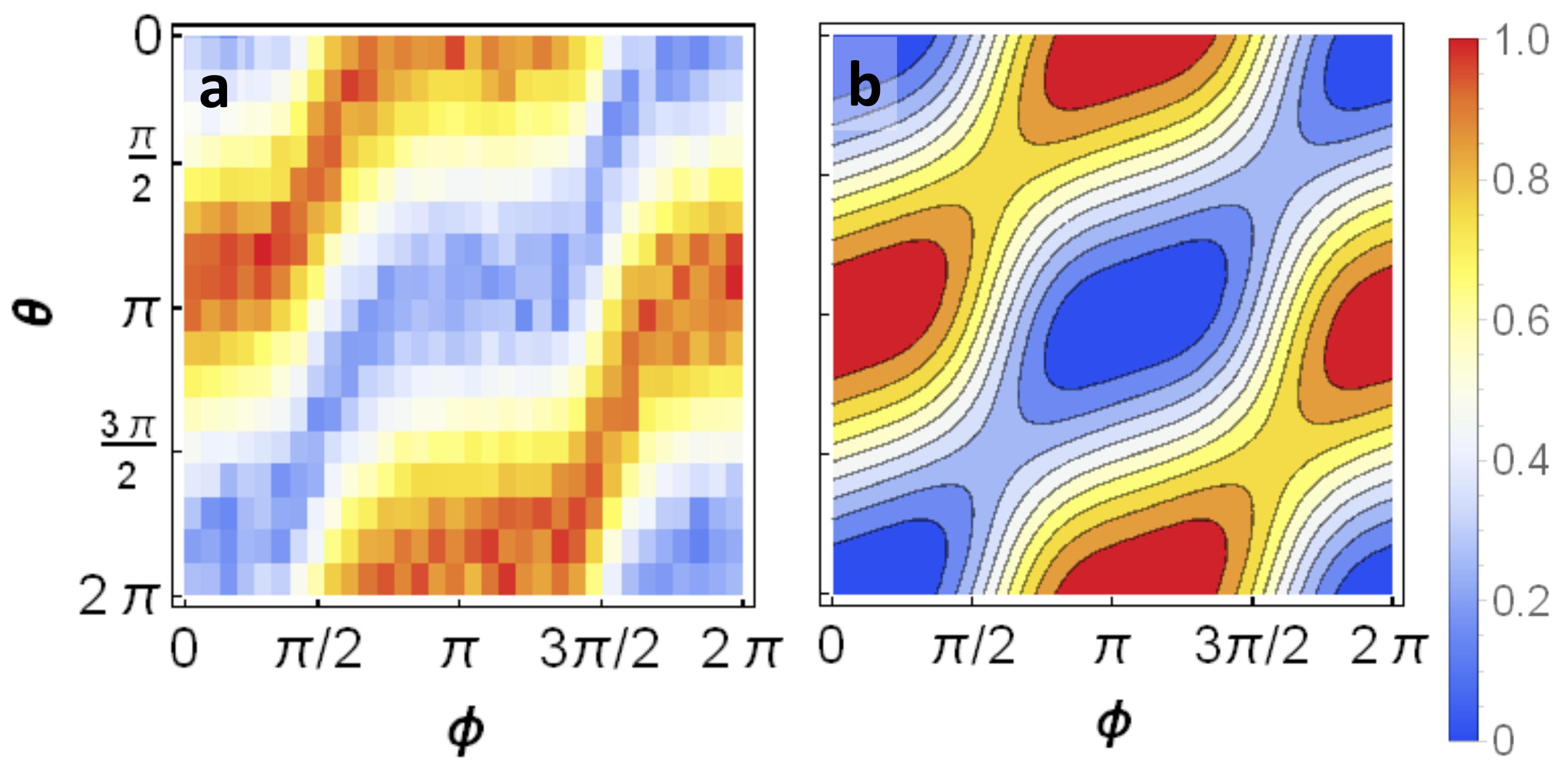}
\caption{\textbf{Experimental data and model of geometrical Ramsey magnetometry. (a)} Data obtained by performing the Ramsey scheme on an ensemble of $^{87}\text{Rb}$ atoms. After the first $\frac{\pi}{2}$-pulse the magnetic field is rotated by an angle $\phi$, then a $\pi$-pulse is used in order to decouple decoherence effect which are due to the trapping, finally a second $\frac{\pi}{2}$-pulse is performed after which the population in the $\ket{2,0}$ state is measured. All three pulse times are calibrated by using the data in Fig. \ref{FigRotScan}, which result in no loss of fringe visibility. \textbf{(b)} Model according to Eq. \eqref{eqnPr20}. The minor-to-major axis ratio has been set to $\tilde{\Omega}=0.27$ in order to fit the data. The measurement range is seen here as fringe visibility is lost around $\phi=\frac{\pi}{2}$.}\label{FigRamsey}\end{figure}

To compare our model predictions to the experiment we implemented the Ramsey sequence described above using a fixed elliptical microwave polarization with two main differences. Firstly, due to technical constraints, instead of scanning the phase of the second $\frac{\pi}{2}$-pulse, we scanned $\phi$ by rotating the magnetic field direction between the two $\frac{\pi}{2}$-pulse. Secondly, we added an additional $\pi$ echo-pulse in between the two $\frac{\pi}{2}$-pulses, in order to mitigate dephasing due to in-homogeneous trap-induced light shifts. All pulse times were calibrated using the Rabi flop data shown in Fig. \ref{FigRotScan} such that the fringe visibility remains constant at all rotation angles, without changing $\theta_f$. By fitting the fringe phase to the expression in Eq. \eqref{eqnThf} we determined the minor-to-major ratio of our polarization ellipse to be $\tilde{\Omega}\approx0.27$. Figure \ref{FigRamsey} shows the theoretical model in Eq. \eqref{eqnPr20} and the measured data, which are in good agreement. The data clearly shows that the clock states population can be manipulated via a magnetic field rotation.

\begin{figure}\includegraphics[width=\columnwidth]{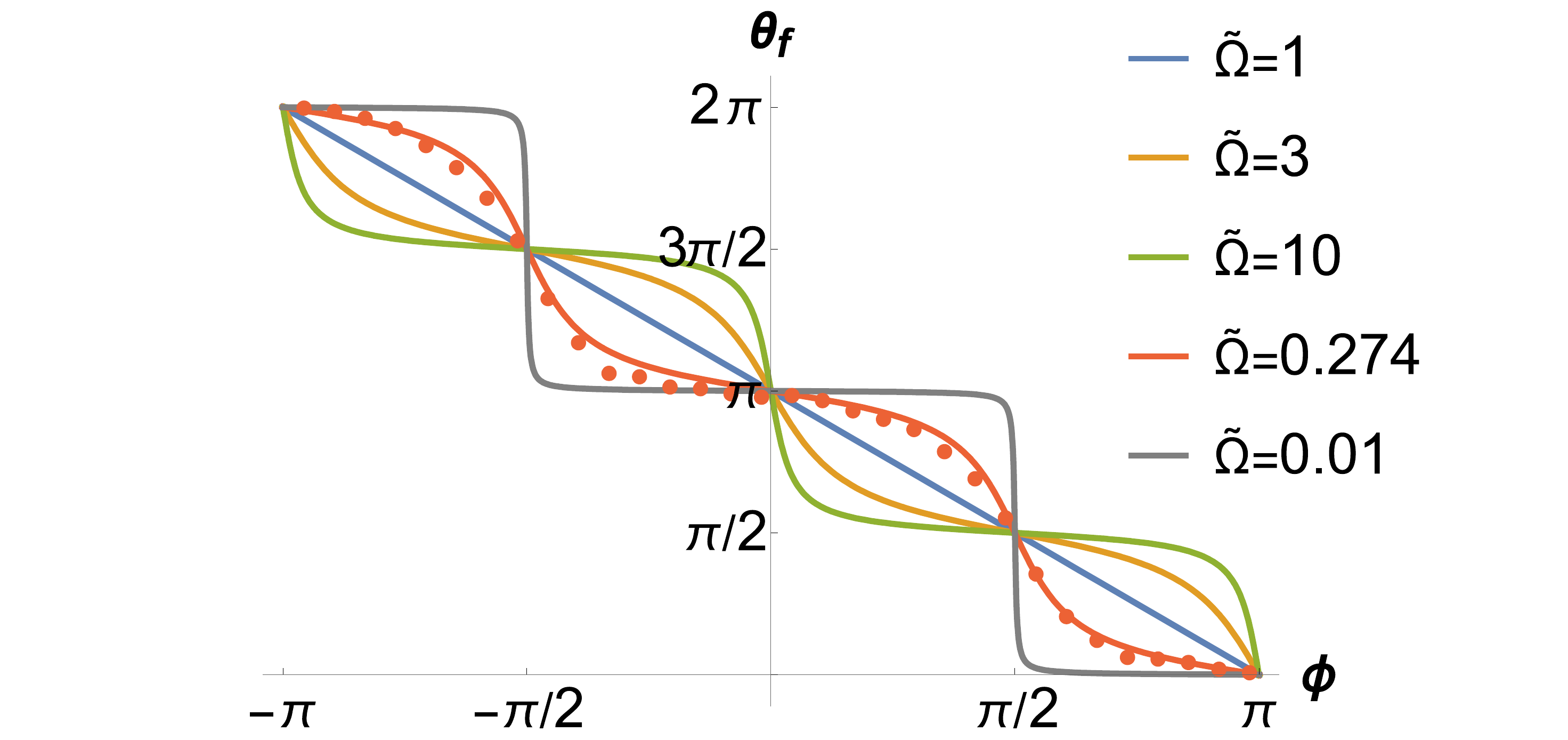}
\caption{\textbf{Geometrical magnetometry by tracking the Ramsey fringe phase.} The Ramsey fringe phase $\theta_{f}$ is the second pulse phase, $\theta$, which maximizes the population of $\ket{2,0}$. Here the trade-off between sensitivity and measurement range, $\delta_{\text{max}}$, is clearly seen as for a minor-to-major ratio $\tilde{\Omega}=1$ (blue) the measurement range is $2\pi$, yet by increasing it the slope around $\phi=0$ increases dramatically but quickly saturates. We superimpose data (red points) obtained from the data shown in Fig. \ref{FigRamsey}a. Fitting these results to the expressions for $\theta_{f}$ we obtain $\tilde{\Omega}\approx0.27$, which shows a good overlap with the theory (red solid). We also highlight the case $\tilde{\Omega}=0.01$ (gray), demonstrating the topological phase of an open trajectory in parameter space described by Robbins and Berry \cite{Robbins1994}.}\label{FigMagDC}\end{figure}

The minor-to-major axis ratio in our scheme determines the measurement sensitivity and range. Figure \ref{FigMagDC} shows the fringe phase for different minor-to-major axis ratios according to Eq. \eqref{eqnThf}. The trade-off between sensitivity and measurement range is evident as larger minor-to-major ratios result in a steeper change in $\theta_{f}$, however it also results in faster flattening of the slope and saturation of sensitivity. Using the experimental data shown in Fig. \ref{FigRamsey} we estimated the fringe phase for each field angle, $\phi$, and added these as data points to Fig. \ref{FigMagDC} (red filled points) on top of the theoretical curve. As seen the data and model are again in good agreement. 

Figure \ref{FigMagDC} also shows the case $\tilde{\Omega}=0.01$ (gray) which corresponds to an almost linear polarization along $\hat{b}$, this setting is obviously entirely insensitive to small field rotations however highlights the topological phase which does not require closing a path in parameter space, described by Robbins and Berry \cite{Robbins1994} and verified experimentally \cite{Usami2007,Takahashi2009,Li2016}, here seen as an abrupt and quantized $\pi$ jump. This phase difference is insensitive to the path the magnetic field orientation takes from $\hat{b}$ to $-\hat{b}$. In this context, by scanning the minor-to-major axes we interpolate between the linear and abrupt phase change, generated by circular and linear polarizations respectively.

\section{VI. AC magnetometry}
A unique property of our geometric method is that the instantaneous magnetic field is sampled at the second $\frac{\pi}{2}$-pulse instant, while Zeeman-Ramsey methods measure a dynamic phase which accumulates in the interval between the two pulses. 

This allows to generalize our magnetometry method to AC signals, by replacing the second $\frac{\pi}{2}$-pulse with a continuous pulse, the amplitude of which is modulated in time to be $\Omega\left(t\right)=\Omega_0\cos\left(\omega_m t\right)$. This intuitively modifies the measurement from an instantaneous single-point sampling to a continuous overlap with the modulating signal, effectively creating a spectral filter.

Explicitly, by assuming a weak oscillating signal magnetic field, causing a rotation of the form $\phi\left(t\right)=\phi_0\cos\left(\omega_0 t\right)$, the probability to find the system in the state $\ket{2,0}$, at $t=\frac{2\pi}{\omega_m}n$ for integer $n$, is given by (see appendix B),
\begin{equation}
    P_2\approx\frac{1}{2}+\phi_{0}\frac{2\Omega_{2}\omega_{0}}{\omega_{0}^{2}-\omega_{m}^{2}}\sin\left(2\pi n\frac{\omega_{0}}{\omega_{m}}\right),\label{eqnFilterAC}
\end{equation}
Equation \eqref{eqnFilterAC} defines a spectral filter, that peaks at $\omega_m\rightarrow\omega_0$ with the value $\frac{1}{2}+2\pi n\phi_{0}\frac{\Omega_{2}}{\omega_m}$ and has width $\Delta\omega=\frac{\omega_m}{2n}$. By scanning $\omega_m$ we obtain the frequency spectrum of the magnetic field. Here we have assumed that the phase of the signal is known and "locked" to the modulation of $\Omega$, however this assumption can be relaxed by performing decoherence-spectroscopy and averaging "unlocked" measurements \cite{Kotler2011,Kotler2013} (see appendix B).

\section{VII. Sensitivity analysis}
We calculate the sensitivity of our magnetometry method assuming only projection noise with no systematic drifts. We approximate the smallest change in $\delta$ that can be observed, $\Delta\delta$, by the Cramer-Rao bound \cite{Rao1945,Cramer1946},
\begin{equation}
    \begin{cases}
        \Delta\delta\approx\frac{1}{\sqrt{N\cdot F\left(\delta=0\right)}}\\
        F\left(\delta\right)=\sum_{\xi=0,1}\mathcal{P}\left(\xi|\delta\right)\left(\frac{d\ln\mathcal{P}\left(\xi|\delta\right)}{d\delta}\right)^{2}
    \end{cases},\label{eqnSensTh}
\end{equation}
where $N$ is the number of independent identically distributed measurements, $F$ is the Fisher information and $\mathcal{P}\left(\xi|\delta\right)$ is a Bernoulli distribution set by the probability to measure the state $\ket{2,0}$, i.e it takes the value $1$ with probability $P_2$, and $0$ otherwise. Since we are concerned with small signals Eq. \eqref{eqnSensTh} is evaluated at $\delta=0$. Using Eq. \eqref{eqnPr20}, with the second $\frac{\pi}{2}$-pulse phase set to $\theta=\frac{\pi}{2}$, we have,
\begin{equation}
    P_{2}=\frac{1}{2}-\frac{1}{2}\sqrt{\frac{\tilde{\Omega}^{2}\delta^{2}}{B_{f}^{2}+\tilde{\Omega}^{2}\delta^{2}}}\sin\left(\frac{\pi}{2}\sqrt{\frac{B_{f}^{2}+\tilde{\Omega}^{2}\delta^{2}}{B_{f}^{2}+\delta^{2}}}\right). \label{eqnPr20Sens}
\end{equation}
Plugging Eq. \eqref{eqnPr20Sens} in Eq. \eqref{eqnSensTh} we obtain the sensitivity,
\begin{equation}
    \Delta\delta_{\varepsilon=0}=B_f\left(\sqrt{N}\tilde{\Omega}\right)^{-1},\label{eqnSens}
\end{equation} 
where the $\varepsilon=0$ subscript indicates this is evaluated in an error-less scenario. As expected the sensitivity improves with more measurements and large minor-to-major ratio. 

Interestingly, the sensitivity in Eq. \eqref{eqnSens} is of geometric origin, i.e the smallest magnetic field rotation, $\phi\sim\frac{\delta}{B_f}$, is inversely proportional to the minor-to-major ratio of the polarization ellipse. The dynamic range is given by the measurement range divided by the sensitivity, yielding $\delta_{\text{max}}/\Delta\delta=\sqrt{N}$ "magnetic" pixels. We note that $B_f$ and $\tilde{\Omega}$ are experimentally controlled, however we expect the sensitivity to be ultimately determined by parameters that are not controlled. 

In theory our sensitivity is unlimited as both $\tilde{\Omega}^{-1}$ and $B_f$ can be arbitrarily reduced. Practically we expect a lower bound on both parameters. Specifically we expect that the clock subspace coherence time, $\tau_\text{clk}$, will limit the attainable sensitivity through $\frac{\hbar}{\mu\tau_\text{clk}}$. In Zeeman-Ramsey based magnetometry methods the measurement is essentially an energy measurement, thus the system's coherence time enters naturally through the energy-time uncertainty relation. As we show below, in our method the coherence time becomes a limitation only when considering errors, namely population "leaks" out of the clock subspace which are caused by the system time-dynamics. Specifically we consider power broadening and coupling to other levels during Rabi nutation between the two clock states, and diabatic transitions during the magnetic field ramp-down.

Starting with power broadening, we note that one cannot improve the accuracy in Eq. \eqref{eqnSens} infinitely by taking the limit $B_f\rightarrow0$ since this will generate unwanted coupling to other Zeeman states outside of the clock subspace (e.g. in the $F=1$ and $F=2$ hyperfine manifolds in $^{87}$Rb). The coupling to nearby states is estimated as $\hbar\left|\hat{b}\times\boldsymbol{\Omega}\right|$, which we bound further by $\hbar\left|\boldsymbol{\Omega}\right|$. Since the RF drive is detuned from other transitions, the population error due to this coupling can be crudely estimated by a power broadening mechanism,
\begin{equation}
    \varepsilon_{\text{P.B}}\approx\frac{1}{1+\frac{\left(\frac{\mu B_{f}}{\hbar}T_{\pi}\right)^{2}}{1+\tilde{\Omega}^{2}}},\label{eqnOm2}  
\end{equation}
where $\varepsilon_{\text{P.B}}$ is the error due to power broadening and we defined the $\pi$-pulse time, $T_\pi=\frac{1}{\Omega_1}$. Eq. \eqref{eqnOm2} shows that minimizing this error requires increasing the product $B_f T_\pi$ as much as possible; i.e. for a fixed $B_f$, slowing down the Rabi frequency. Here, the system coherence time, $\tau_\text{clk}$, enters as an upper bound to $T_\pi$ .

A second dynamical error is that caused by diabatic transitions which occur as the magnetic field is decreased, during a ramp time $T$, from $B\left(0\right)=B_i$ to  $B\left(T\right)=B_f$. These transitions may again lead to transitions to states outside of the clock subspace and thus to a loss of measurement accuracy. To capture these effects we consider the interaction Hamiltonian between one of the clock states and a nearby Zeeman state with $m_F=\pm1$ (other $m_F$ states are uncoupled). This has the form $H_{d}=m_F\mu_F\left(\frac{B\left(t\right)}{2}\left(\sigma_{z}+\mathbb{I}\right)+\delta\sigma_{x}\right)$, where $\mu_F$ is a ($F$ dependent) magneton coupling and we assumed, without loss of generality, that $\delta$ is directed towards the $\hat{x}$ direction. This is of course the renowned Landau-Zener-Stuckelberg (LZS) Hamiltonian. For a linear $B\left(t\right)=\alpha t$ it is exactly solvable \cite{Landau1932,Zener1932,Stueckelberg1932}. Traditionally the system is initialized in the lower band at $B\rightarrow-\infty$ and driven through the gap at $B=0$, to $B\rightarrow\infty$. Landau, Zener and Stuckelberg calculated exactly the diabatic probability, i.e the probability to find the system in the upper band. Here the two bands are the clock-state and Zeeman state respectively.

We need to consider a variation of the LZS problem. We initialize the system on the lower band at a finite $B=B_i$ and drive the system to $B=B_f$ without ever crossing the gap. We are interested in the probability of diabatic transitions to the upper band, $\varepsilon_{\text{D}}$, in terms of the instantaneous Hamiltonian basis, as opposed to in the non-interacting basis, at $B\rightarrow\pm\infty$, and as a function of the ramp-down time, $T$, and the ramp profile. Of course, for a linear ramp profile, one can relate the full solution of LZS to the problem presented here. However, as we show below, we are interested in more general ramp profiles, as a linear profile is far from optimal. 

Hence we work in the instantaneous eigenbasis, with the Hamiltonian, $H_\text{inst}=\Delta E\left(t\right)\sigma^z+\hbar\dot{\gamma}\sigma^x$, where $\Delta E$ is the time-dependent energy difference between the clock-state and Zeeman state, and $\gamma=\frac{1}{2}\arctan\left(\frac{2\delta}{B}\right)$. We note that $\gamma$ approaches the magnetic field angle, $\phi$, as $\frac{\delta}{B}\rightarrow0$. By changing to an interaction picture with respect to the diagonal part and using a first order Dyson series we approximate the unitary evolution operator (see appendix C), and obtain the estimation,
\begin{equation}
    \varepsilon_{D}\approx\left|\int_{0}^{T}dt\dot{\gamma}e^{\frac{i}{\hbar}\int_{0}^{t}dt^{\prime}\Delta E\left(t^{\prime}\right)}\right|^{2}.\label{eqnEstDiabProb}
\end{equation}

For a linearly varying $\gamma$ (which is approximately a linearly varying magnetic field angle), Eq. \eqref{eqnEstDiabProb} can be exactly solved and bound by, 
\begin{equation}
    \varepsilon_{\text{D}}\left(\delta\right)\approx\frac{\left(\frac{\delta}{B_{f}}\right)^{2}}{1+\left(\frac{\mu B_{f}}{\hbar}\cdot T\right)^{2}}\leq\frac{\left(\frac{\delta}{B_{f}}\right)^{2}}{1+\frac{1}{2}\left(\frac{\mu B_{f}}{\hbar}\cdot T\right)^{2}},\label{eqnRamp} 
\end{equation}
where the latter bound is used in order to make the approximation stringent. We note that by assuming that $T\sim\left(\frac{\mu\delta}{\hbar}\right)^{-1}$ we may approximate the ramp time as $\frac{T}{\tau_{\delta}}\approx\sqrt{\frac{1}{\varepsilon_{\text{D}}}}\cdot\left(\frac{\delta}{B_{f}}\right)^{2}$, which is written in units of the characteristic period of the Zeeman frequency shift due to the field $\delta$, $\tau_\delta=\left(\frac{\mu\delta}{\hbar}\right)^{-1}$. This characteristic time-scale, $\tau_\delta$, is also the relevant time-scale for Zeeman based magnetometry methods as it is the typical time required to accumulate a substantial phase due to the signal. As an example, for the scenario, $B_f=5\delta$ and $\varepsilon_{\text{D}}=0.01$, we obtain $T<\tau_\delta$, i.e a ramp time that is shorter than a typical Ramsey experiment. 

We note that even for an abrupt field ramp, i.e $T\rightarrow0$, which will be important below, our approximation in Eq. \eqref{eqnRamp} remains valid, as can be seen by a direct calculation.

Assuming a finite system coherence time, $\tau_\text{clk}$, we would like to choose the parameters $T$, $T_\pi$ and $\tilde{\Omega}$, under the constraint $T+T_\pi=\tau_\text{clk}$, such that the sensitivity is optimized (minimized). The resulting relations between these parameters and $\tau_\text{clk}$ then yields a sensitivity which depends, in leading order, on the system coherence time.

For a generic "leak" error out of the clock subspace by an amount $\varepsilon$ we obtain a more refined version of Eq. \eqref{eqnSens}, namely, $\Delta\delta=\Delta\delta_{\varepsilon=0}\sqrt\frac{1+\varepsilon}{1-\varepsilon}=\Delta\delta_{\varepsilon=0}\left(1+\varepsilon\right)+\ord{\varepsilon^2}$. We may now use the error bounds in the right-hand side of Eq. \eqref{eqnOm2} and \eqref{eqnRamp} in order to incorporate power broadening and diabatic transition errors in an independent way,
\begin{equation}
    \Delta\tilde{\delta}=\frac{1}{\sqrt{N}}\frac{\tilde{B}_{f}}{\tilde{\Omega}}\cdot\sqrt{\frac{1+\varepsilon_{\text{D}}\left(\delta_\text{max}\right)}{1-\varepsilon_{\text{D}}\left(\delta_\text{max}\right)}}\cdot\left(\sqrt{\frac{1+\varepsilon_{\text{P.B}}}{1-\varepsilon_{\text{P.B}}}}\right)^{2},\label{eqnSensFull}
\end{equation}
where we defined a dimensionless magnetic field $\tilde{B}=\frac{\mu B}{\hbar}\tau_\text{clk}$, in terms of its corresponding Zeeman splitting frequency scaled by the coherence times, and use the measurement range, $\delta_\text{max}$ in the expression for the diabatic transition error in Eq. \eqref{eqnRamp}. The power broadening error, $\varepsilon_\text{P.B}$, in \eqref{eqnSensFull} is squared in order to independently account for two pulses. 

Both corrections in Eq. \eqref{eqnSensFull} (second and third terms) are minimized as the final magnetic field, $\tilde{B}_f$, is taken to be arbitrarily large, yet the leading-order sensitivity (first term) is linear in it. An optimum to the sensitivity therefore can be found in the $\tilde{\Omega},\tilde{B}_f$ plane. To simplify our analysis we assume a linear relation, $\tilde{\Omega}\propto\tilde{B}_f$. With this assumption we may safely expand Eq. \eqref{eqnSensFull} in large $\tilde{B}_f$ and minimize the resulting sensitivity yielding,
\begin{equation}
    \Delta\tilde{\delta}=\frac{2\sqrt{2}}{\sqrt{N}\left(1-\tilde{T}\right)}\left(1+\frac{1}{\left(1-\tilde{T}\right)^{2}\tilde{B}_{f}^{2}}\right)+\ord{\tilde{B}_{f}^{-4}}, \label{eqnSensExp}
\end{equation}
where we used the dimensionsless time $\tilde{T}=T/\tau_\text{clk}$. Indeed Eq. \eqref{eqnSensExp}, is obtained by the optimal linear relation, $\tilde{\Omega}=\frac{1-\tilde{T}}{\sqrt{2}}\tilde{B}_{f}$.

The sensitivity in Eq. \eqref{eqnSensExp} is optimized by using an abrupt magnetic field ramp, $\tilde{T}\rightarrow0$. This yields the sensitivity,
\begin{equation}
    \frac{\mu\Delta\delta}{\hbar}=\frac{2\sqrt{2}}{\sqrt{N}\tau_\text{clk}}\left(1+\tilde{T}+\frac{1+3\tilde{T}}{\tilde{B}_{f}^{2}}\right)+\ord{\tilde{T}^2,\tilde{B}_{f}^{-4}},\label{eqnSensSol}
\end{equation}
where we have retained linear correction in $\tilde{T}$, accounting for practical minimal field ramp times. As expected, the sensitivity in Eq. \eqref{eqnSensSol} is inversely dependent on the clock subspace coherence time, $\tau_\text{clk}$.

The factor $2\sqrt{2}$ in Eq. \eqref{eqnSensSol} is explained by the linear relation between $\tilde{\Omega}$ and $\tilde{B}_{f}$, which enters in the leading order sensitivity, $\frac{\tilde{B}_{f}}{\tilde{\Omega}}$ and in the correction due to power broadening, $\left(1+\varepsilon_{\text{P.B}}\right)\left({1-\varepsilon_{\text{P.B}}}\right)^{-1}$, in Eq. \eqref{eqnSensFull}. Interestingly, even in this abrupt field ramp regime the corrections due to diabatic transitions do not enter in leading-order. To ensure our derivations are valid we have investigated the sensitivity numerically and in a self-consistent manner, yielding the same results as above (see appendix D).

The sensitivity in Eq. \eqref{eqnSensSol} can be converted to the more conventional atomic magnetometry form with the identification $N=n\cdot V\cdot \frac{T_\text{total}}{\tau_\text{clk}}$, where $n$ is the atomic ensemble density, occupying a volume $V$ and $\frac{T_\text{total}}{\tau_\text{clk}}$ is the number of repetitions, given by the ratio between the total measurement time, $T_\text{total}$, and the coherence time. In this approach, the total measurement time is kept fixed and the sensitivity scales as $\frac{1}{\sqrt{\tau_\text{clk}}}$.

We note that the coherence time enters similarly in the AC magnetometry method, through the stroboscopic time, $n=\frac{\omega_m}{2\pi}t$, appearing in Eq. \eqref{eqnFilterAC}, such that longer measurement times generate a narrower spectral filter with a higher peak. Similar to the DC case, the coherence time bounds the maximal measurement time, thus setting a limit on the spectral resolution and measurement sensitivity.

Finally, the sensitivity is optimized by setting the largest possible $\Omega_2$ and the smallest possible ramp time $T$. The remaining parameters are then given by,
\begin{equation}
    \begin{cases}
    \frac{\mu B_{f}}{\hbar}=\sqrt{2}\Omega_{2}\\
    \Omega_{1}=\frac{1}{\tau-T}
    \end{cases},\label{eqnSettings}
\end{equation}
which enforces the relations between the three system parameters as discussed above (next-order correction to Eq. \eqref{eqnSettings} appear in appendix D).

We compare our proposed geometric method to the more conventional Zeeman-Ramsey method, i.e two Zeeman-split quantum states are used in order to measure a small magnetic field $\delta_z$, which is parallel to a set quantization field, $B_z$. 

The Zeeman-Ramsey measurement is performed by using an AC field to employ a $\frac{\pi}{2}$-pulse, which creates an equal superposition of two Zeeman-split states. After a wait time $T$ the superposition acquires a differential phase $\Delta\omega T$, where $\Delta\omega$ is the frequency difference between the AC field frequency and the frequency due to the splitting between the two states. By tuning the AC field frequency to $\frac{\mu B_z}{\hbar}$ then $\Delta\omega=\frac{\mu\delta}{\hbar}$. A second $\frac{\pi}{2}$-pulse is then used to "close" the superposition. The population in the excited state is now given by
\begin{equation}
    P_{\text{Ramsey}}=\frac{1}{2}+\frac{1}{2}\cos\left(\frac{\mu\delta}{\hbar}T\right),\label{eqnPopRamsey}
\end{equation}
Assuming $T$ is limited by some coherence time, $\tau_\text{Z}$, we may evaluate Eq. \eqref{eqnSensTh} to obtain $\Delta\delta_{\varepsilon=0}=\frac{2}{\sqrt{N}\tau_\text{Z}}$. Clearly this sensitivity and the sensitivity derived in Eq. \eqref{eqnSensSol} scale similarly, showing that the Zeeman-Ramsey and the geometrical magnetoemtry methods are comparable. 

However, coherence times in a clock subspace are typically much longer than in a Zeeman-split subspace \cite{Langer2005,Kleine2011}. This implies that the proposed geometric magnetometry method may improve upon the sensitivity of Zeeman-splitting based magnetometry methods. 

For our geometric method, a systematic error or a fluctuation of the quantization field, $B_f\rightarrow B_f+\Delta B_{f,\perp}$, results in a measurement error, $\delta\rightarrow\delta\left(1+\frac{\Delta B_{f,\perp}}{B_f}\right)$. That is, even an error of magnitude $\Delta B_{f,\perp}\sim\delta$ will not effect the measurement result significantly as $\delta/B_f$ is assumed to be small. Hence, our method is insensitive to errors in the quantization field. However any stray field, $\Delta B_{f,\parallel}$, that is parallel to $\delta$ (e.g created by coils that generate $B_f$), cannot be distinguished from $\delta$. Such errors need to be added to the expression in Eq. \eqref{eqnSensFull}.

In like manner, a Zeeman based method cannot distinguish between noise in the quantization field, $B_z$, and signal, $\delta_z$, and is insensitive in leading order to any noise perpendicular to the quantization axis. That is both methods are insensitive to magnetic field noise in one direction and cannot distinguish between error and signal in the perpendicular direction. However field noises along the quantization axis are inherent to experimental implementations and cannot be easily overcome.

\section{VIII. Conclusions}
We introduced in this paper a magnetometry method which employs clock states, i.e quantum states whose energy is independent of the magnetic field. We showed that even though there is no dynamical phase that depends on the magnetic field that we aim to measure, the orientation of the wave function can still be used to measure magnetic fields which are perpendicular to the quantization magnetic field. We employed this concept to propose both DC and AC magnetometry methods.

Our method's sensitivity, i.e the smallest measurable magnetic field, is $\Delta\delta=\frac{\hbar}{\mu\tau_{\text{clk}}}\sqrt{\frac{8}{N}}$ 
in the high-field limit, where $N$ is the number of independent measurements and $\tau_{\text{clk}}$ is the clock states subspace coherence time. A sensitivity that scales inversely with the coherence time is common also in Zeeman based magnetometry methods, however the coherence time in a clock subspace is typically much larger than that of a Zeeman-split subspace. This implies that a  magnetometry method based on clock states may improve upon the sensitivity of contemporary magnetometers. 

Finally, we have demonstrated a few of our proposed methods on an ensemble of trapped $^{87}\text{Rb}$ atoms, which display an excellent agreement with our derivations.

\begin{acknowledgments}
This work was supported by the Crown Photonics Center, ICore-Israeli excellence center circle of light, the Israeli Science Foundation, the Israeli Ministry of Science Technology and Space, the Minerva Stiftung and the European Research Council (consolidator grant 616919-Ionology)
\end{acknowledgments}

\section{Appendix A: Two spin-$\frac{1}{2}$ in the lab frame}

In the main text we show that the degenerate $\ket{S}$ and $\ket{T}$ states, which make up the clock subspace, are used for magnetometry. Our derivations are performed in the "magnetic" frame, in which the triplet state is invariant under rotations of the magnetic field (the singlet is trivially invariant as well). Here we repeat this derivation in the static "lab" frame, which remains constant and is written in the $\sigma_1^z+\sigma_2^z$ eigenstates basis.

The system Hamiltonian, given in Eq. \eqref{eqnHam} of the main text, is,
\begin{equation}
H=\mu\boldsymbol{B}\left(t\right)\cdot\left(\boldsymbol{\sigma}_{1}+\boldsymbol{\sigma}_{2}\right)+\hbar\boldsymbol{\Omega}\left(t\right)\cdot\boldsymbol{\sigma}_{1}. \label{eqnHamSupp}
\end{equation}

The magnetoemetry sequence is performed by initializing the system in the $\ket{S}$ state with a large quantization field pointing to the $\hat{z}$ direction and performing an adiabatic $\frac{\pi}{2}$-pulse. The pulse acts exclusively in the clock subspace and creates a superposition of the two clock states. The magnetic field magnitude is then reduced and the field is rotated by an angle $\chi+\frac{\pi}{2}+\phi$, where $\chi$ is the angle of $\boldsymbol{\Omega}$ with respect to the $\hat{z}$ direction and $\phi$ is an additional uncontrolled rotation due to the signal magnetic field, $\delta$ (see Fig. \ref{FigTwoSpins} of the main text). A second adiabatic $\frac{\pi}{2}$-pulse then encodes population in the $\ket{S}$ state which is linear in $\phi$, to leading order.

The two adiabatic pulses are implemented via the second, $\boldsymbol{\Omega}\left(t\right)$-dependent term in Eq. \eqref{eqnHamSupp}, where the time-dependence is used for both switching the pulse on and off and ramping its amplitude adiabatically. To simplify this term we assume the magnetic field lies on the $\hat{x}-\hat{z}$ plane at an angle $\alpha$ with the $\hat{z}$ direction. We then rewrite the pulse term as,
\begin{equation}
    \begin{cases}
    V\left(\alpha\right)=P\left(\alpha\right)\left[\hbar\Omega\left(\cos\left(\chi\right)\sigma_{1}^{z}+\sin\left(\chi\right)\sigma_{1}^{x}\right)\right]P\left(\alpha\right)\\
    P\left(\alpha\right)=\ketbra SS+e^{-i\alpha\hat{J}_{y}}\ketbra{1;0}{1;0}e^{i\alpha\hat{J}_{y}},\label{eqnAdiaPulse}
    \end{cases}
\end{equation}
where $\Omega T$ corresponds to the pulse area and $P\left(\alpha\right)$ is a projection on the instantaneous singlet and triplet states, such that the pulse does not allow population to "leak" outside of the clock subspace. As prescribed in the main text, we will set $\Omega T\cos\left(\chi\right)=\frac{\pi}{2}$. 

\begin{figure}[h]\includegraphics[width=\columnwidth]{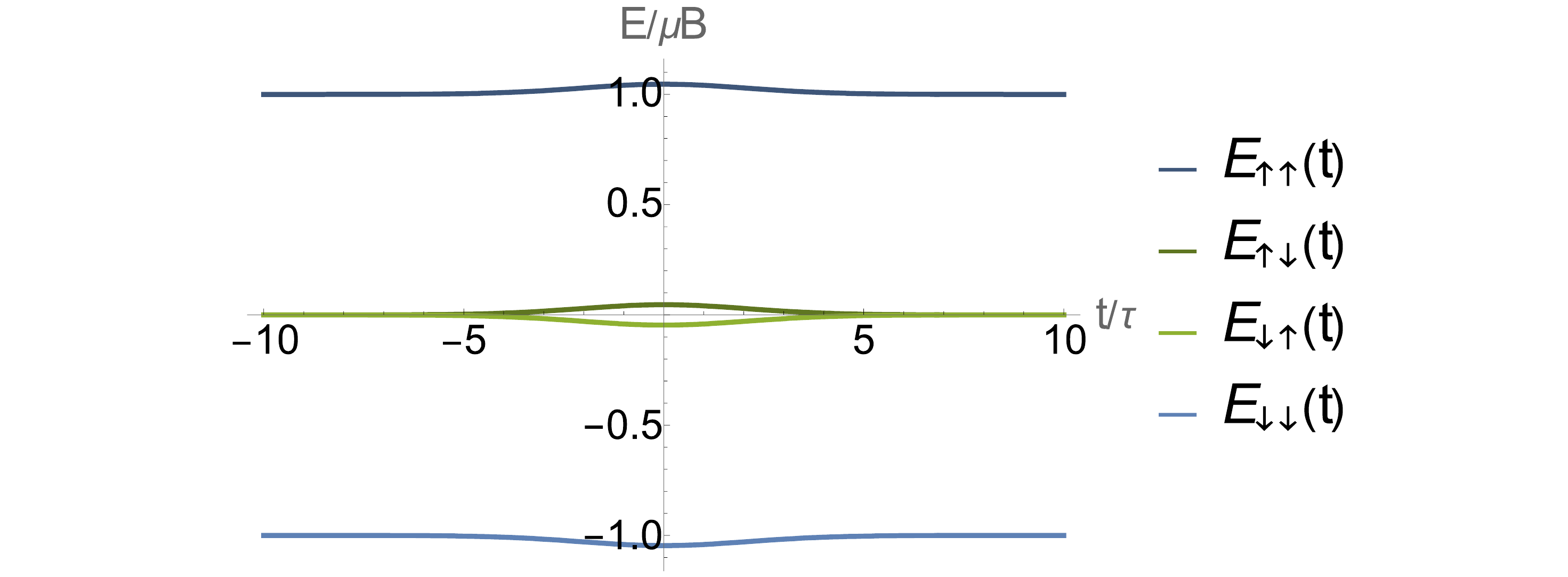}
\caption{\textbf{Two spin-$\frac{1}{2}$ system energy levels during adiabatic pulse.} The state's spin-labels are exact at $t\rightarrow\pm\infty$ and approximate during the pulse. In this example the adiabatic pulse is implemented such that $\Omega\left(t\right)$ is an ideal Gaussian that is ramped from $t=-\infty$ to $t=\infty$. The Gaussian's width is $\tau=\frac{25\sqrt{2\pi}}{\mu B \cos\left(\chi\right)}$, which is set such that its area is $\frac{\pi}{2\cos\left(\chi\right)}$ and peak height is $0.01\mu B$. With these settings the maximal mixing between the $\ket{\uparrow\downarrow}$ ($\ket{\downarrow\uparrow}$) state at $t=0$ with the $\ket{\downarrow\downarrow}$ ($\ket{\uparrow\uparrow}$) state at $t\rightarrow-\infty$ is at most $1\%$. The figure also shows pictorially how the pulse works, i.e the degenerate clock subspace is split to the $\ket{\uparrow\downarrow}$ and $\ket{\downarrow\uparrow}$ states, which then accumulates a $\frac{\pi}{2}$ phase difference, creating the singlet and triplet superposition.}\label{figLevelSplitting}\end{figure}

While we do not rigorously prove the validity of Eq. \eqref{eqnAdiaPulse}, Fig. \ref{figLevelSplitting} shows an example of the system instantaneous spectrum, as a function of time, while an ideal adiabatic $\frac{\pi}{2}$-pulse is applied. Clearly the degenerate clock subspace (green lines) is split, but remains at relatively large separation from the other states (blue lines). We also refer the reader to Appendix C, in which diabatic transitions are discussed more rigorously, in a different but analogous context.

The form of Eq. \eqref{eqnAdiaPulse} allows to write down the entire sequence more easily. It is,
\begin{equation}
    \text{Pr}\left(S\right)=\left|\broket S{e^{iV\left(\chi+\frac{\pi}{2}+\phi\right)T}e^{-i\left(\chi+\frac{\pi}{2}+\phi\right)\hat{J}_{y}}e^{iV\left(0\right)T}}S\right|^{2},\label{eqnProbLab}
\end{equation}
where the first (right-most) exponential accounts for the first pulse, the second (middle) for the magnetic field rotation and the last (left-most) for the second pulse. Computing this expression explicitly recovers exactly Eq. \eqref{eqnPrSpinFull} of the main text. 

We note that in Eq. \eqref{eqnProbLab} above we explicitly excluded free evolution due to the $\boldsymbol{B}\left(t\right)$-dependent term of the Hamiltonian in Eq. \eqref{eqnHamSupp}. This term cannot affect our result since it does not act in the clock subspace in which all the manipulations occur.

\section{Appendix B: AC magnetometry}
Similar to the DC magnetometry scheme, we start by initializing the system at $\ket{2,0}$ with the magnetic field set to $B=B_i\gg\delta$. We then preform a $\frac{\pi}{2}$-pulse, followed by lowering the magnetic field adiabatically to the value $B_f$. Then, We turn on the amplitude-modulated RF-drive continuously, such that the Hamiltonian is given by Eq. \eqref{eqnHamClk}, with the modification $\boldsymbol{\Omega}\rightarrow\boldsymbol{\Omega}\cos\left(\omega_m t\right)$. This is no longer a two-pulse problem, rather, a hard time-continuous problem in which the Hamiltonian does not commute with itself at different times.

To overcome this we assume a small signal $\delta$, and therefore a small field rotation angle $\phi\left(t\right)$, which we can solve in linear order in  $\phi$. This simplifies Eq. \eqref{eqnHamClk} to
\begin{equation}
    H=\frac{\hbar\Omega_{1}}{2}\cos\left(\omega_{m}t\right)\left(\tau^{y}-\tilde{\Omega}\phi\tau^{x}\right)+\ord{\phi^2}.\label{eqnHamAC}
\end{equation}

Thus we approximate the resulting unitary evolution operator as $U\approx e^{-\frac{i}{\hbar}\int\limits _{0}^{t}H\left(\tau\right)d\tau}$. We note that this is simply the leading-order term in a Dyson expansion, the next order term contributes only an additional $\tau^z$ rotation and vanishes when the modulation is on resonance with $\phi\left(t\right)$.

Assuming a spectral decomposition $\phi\left(t\right)=\int\limits _{-\infty}^{\infty}d\omega\left|\phi\left(\omega\right)\right|e^{i\alpha\left(\omega\right)}e^{i\omega t}$, with the amplitude $\left|\phi\left(\omega\right)\right|$ and phase $\alpha\left(\omega\right)$, we obtain the approximate unitary evolution operator,
\begin{equation}
    \begin{cases}
        U=e^{i\frac{\Omega_{2}}{\omega_{m}}\tau^{x}\int\limits _{0}^{\infty}d\omega\left|\phi\left(\omega\right)\right|F\left(\omega\right)}\\
        F\left(\omega\right)=\frac{2\omega\omega_{m}}{\omega^{2}-\omega_{m}^{2}}\left(\sin\left(\alpha\left(\omega\right)+2\pi n\frac{\omega}{\omega_{m}}\right)-\sin\left(\alpha\left(\omega\right)\right)\right)
    \end{cases},\label{eqnUAC}
\end{equation}
where we used the stroboscopic time $t=\frac{2\pi}{\omega_m}n$ with $n\in\mathbb{Z}$. Since this is a linear-order approximation we might as well assume a monotone, $\left|\phi\left(\omega\right)\right|=\phi_{0}\delta\left(\omega-\omega_{0}\right)$. For simplicity we further assume $\alpha=0$, i.e the phase of $\phi$ is known, this assumption is later relaxed. We obtain the probability,
\begin{equation}
    P_2\approx\frac{1}{2}+2\phi_{0}\frac{\Omega_{2}\omega_{0}}{\omega_{0}^{2}-\omega_{m}^{2}}\sin\left(2\pi n\frac{\omega_{0}}{\omega_{m}}\right),\label{eqnFilterACSup}
\end{equation}
corresponding to Eq. \eqref{eqnFilterAC} of the main text. Equation \eqref{eqnFilterACSup} defines a spectral filter, that peaks at $\omega_m\rightarrow\omega_0$ with the value $\frac{1}{2}+2\pi n\phi_{0}\frac{\Omega_{2}}{\omega_m}$ and has a width $\Delta\omega=\frac{\omega_m}{2n}$. By scanning $\omega_m$ a spectrometer-like scan of the field $\delta$ is made possible. 

\begin{figure}\includegraphics[width=\columnwidth]{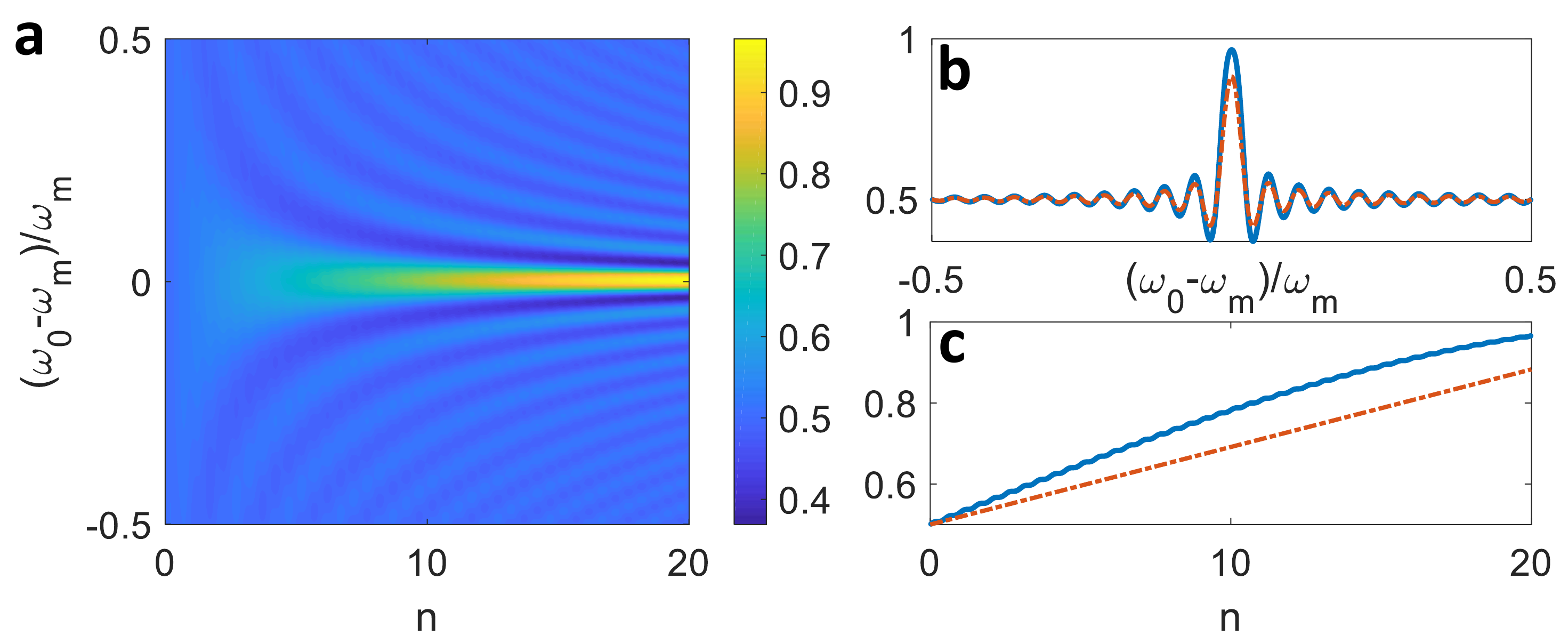}
\caption{\textbf{AC magnetometry, comparing simulation to approximation. (a)} Simulation of population in the $\ket{2,0}$ state (color) for different monotone frequencies $\omega_0$ and stroboscopic times $n$. Each horizontal line shows the population evolution of the $\ket{2,0}$ state due to a magnetic field angle rotating at frequency $\omega_0$ with amplitude $\phi_0=0.005\text{ rad}$. This is detected with a amplitude-modulated RF drive at frequency $\omega_m$ and amplitude $\Omega_{1}=\frac{\omega_{m}}{n}\cdot\frac{1}{3\phi_{0}\tilde{\Omega}}$. \textbf{(b)} Vertical cross section of (a) at $n=20$ showing the spectral filter formed by the continuous amplitude-modulated drive (solid blue). The filter peaks at $\omega_0=\omega_m$ and is in a very good agreement with the approximation (dashed red). \textbf{(c)} Cross section of (a) at $\omega_0=\omega_m$ showing the build-up of population (solid blue) and linear approximation (dashed red) according to Eq. \eqref{eqnFilterAC}. The population exhibits oscillations which peak at the stroboscopic times, emphasizing the importance of interrogating the system at integer periods of the modulation frequency.}\label{FigAC}\end{figure}

To verify our linear-order approximation we compare these results to a Schrodinger's equation time-step simulation. Figure \ref{FigAC}a shows the probability of measuring the system in the $\ket{2,0}$ state as a function of different interrogation times, $n$, and signal frequencies, $\omega_0$. The physical parameters for the simulation are chosen such that a small signal, $\phi_0=0.005\text{ rad}$, is detected with $n=20$ modulation periods and with $\tilde{\Omega}=3$. To keep the measured signal within the linear approximation the drive amplitude is set to $\Omega_{1}=\frac{\omega_{m}}{n}\cdot\frac{1}{3\phi_{0}\tilde{\Omega}}$, corresponding to a maximal rotation of $\frac{\pi}{6}$ radians on the Bloch sphere. Figure \ref{FigAC}b shows a vertical cross section of the simulation at $n=20$ in comparison to the linear approximation in Eq. \eqref{eqnFilterAC}. The spectral filter is evident as the systems response is narrow and peaks at $\omega_m=\omega_0$, both for the simulated results (solid) and the approximation (dashed). Furthermore, we note the importance of using stereoscopic times, as the population peaks at half-integer periods of the modulation frequency as seen in the horizontal cross section at $\omega_m=\omega_0$ in Fig. \ref{FigAC}c. 

This AC magnetometry method intuitively works since our pulses "sample" the magnetic field at different times and overlap it with the modulation signal. For enough modulation periods the overlap approaches an orthogonal projection on the spectral component $\left|\phi\left(\omega\right)\right|=\phi_{0}\delta\left(\omega-\omega_{0}\right)$. This is conceptually different from Ramsey-based AC magnetometry methods which use a train of $\pi$-pulses at a modulation frequency in order to periodically flip the direction in which the dynamical phase is accumulated.

So far we assumed the phase of oscillations, $\alpha$, is "locked" to the modulation. This assumption is conventionally relaxed in one of two methods. The phase of modulation, $\alpha_m$ may be scanned as well, which allows to measure both amplitude and phase information of $\delta$, however this requires the different experimental repetitions to be phase-locked to each other, moreover the phase information of the signal is typically useless. The second method is to intentionally keep the modulation signal unlocked between repetitions. This acts to randomize $\alpha$ at each repetition, by taking the average deviation from $0.5$ of each of the $N$ measurement squared, we obtain the spectrometer filter
\begin{equation}
    \frac{1}{N}\sqrt{\sum\left(p_{i}-\frac{1}{2}\right)^{2}}\rightarrow2\sqrt{2}\phi_{0}\left|\frac{\Omega_{2}\omega\sin\left(\pi n\frac{\omega}{\omega_{m}}\right)}{\omega^{2}-\omega_{m}^{2}}\right|,\label{eqnSpectrometer}
\end{equation}
where the convergence is for a large ensemble of measurements, $N$, assuming uniform sampling of $\alpha\in\left[0,2\pi\right)$. Eq. \eqref{eqnSpectrometer} defines a phase-independent spectral filter that peaks at $\omega_m\rightarrow\omega_0$ with the value $\sqrt{2}\pi n\frac{\Omega_{2}}{\omega_{m}}\phi_{0}$, and with width $\Delta\omega=\frac{\omega_m}{n}$.

\section{Appendix C: Diabatic transition probability}
As described in the main text, the minimal measurement time is constrained by diabatic transitions which occur as the magnetic field is ramped down from $B\left(0\right)=B_i$ to  $B\left(T\right)=B_f$. These transitions may lead to excitations outside of the clock subspace and thus to a loss of measurement accuracy. To capture these effects we consider the interaction Hamiltonian between one of the clock states and a nearby Zeeman state. This has the form
\begin{equation}
    H_{\text{D}}\left(t\right)=\frac{\mu B\left(t\right)}{2}\left(\sigma_{z}+\mathbb{I}\right)+\mu\delta\sigma_{x},\label{eqnDiabaticLab}
\end{equation}
where $\mu$ is a ($F$ and $m_F$ dependent) magnaton coupling formed by a linear combination of $\mu_J$ and $\mu_I$ and we assumed, without loss of generality, that $\delta$ is directed towards the $\hat{x}$ direction. The instantaneous Hamiltonian in Eq. \eqref{eqnDiabaticLab} is diagonlalized by
\begin{equation}
    \begin{cases}
     H_{\text{D}}=RDR^{-1}\\
     R=\cos\left(\gamma\right)\mathbb{I}-i\sin\left(\gamma\right)\sigma^y\\
     D=\text{diag}\left(\frac{\mu B\left(t\right)}{2}\pm\sqrt{\left(\frac{\mu B\left(t\right)}{2}\right)^{2}+\left(\mu\delta\right)^{2}}\right)
     \end{cases},\label{eqnDiabDiag}
\end{equation}
with $\tan\left(2\gamma\right)=\frac{2\delta}{B\left(t\right)}$. Rewriting Schrodinger's equation in the instantaneous eigenbasis set by Eq. \eqref{eqnDiabDiag} we obtain the Hamiltonian, $H_{\text{D,inst}}=\frac{1}{2}\Delta E\left(t\right)\sigma_{z}+\hbar\dot{\gamma}\sigma_{y}$, with $\Delta E=\mu\sqrt{B^{2}\left(t\right)+\left(2\delta\right)^{2}}$. 

It seems that this has only complicated things since now all the matrix elements of the Hamiltonian are time dependent. However this approach has several advantages compared with the lab-frame Hamiltonian (and the LZS formal solution). It is readily written in the instantaneous eigenbasis so by initializing $\ket{\psi\left(t=0\right)}=\ket{\uparrow\left(t=0\right)}$, our sought diabatic transition probability is $\varepsilon_{D}=\left|\broket{\downarrow\left(T\right)}{U\left(T;0\right)}{\up\left(t=0\right)}\right|^{2}$. It is also written in terms of $\gamma$, which for $\delta\ll B$ approaches the magnetic field angle, $\phi$, yielding intuitive and geometric equations (note however that for $B=0$ we have $\gamma=\frac{\pi}{4}$ and not $\frac{\pi}{2}$). 

Changing to an interaction picture with respect to the diagonal term we obtain the interaction Hamiltonian
\begin{equation}
    \begin{cases}
    H_{\text{D,I}}=\hbar\dot{\gamma}\left(\cos\left(\xi\right)\sigma_{y}+\sin\left(\xi\right)\sigma_{x}\right)\\
    \xi=\frac{1}{\hbar}\int_{t_{i}}^{t}dt^{\prime}\Delta E\left(t^{\prime}\right)
    \end{cases}.\label{eqnDiabInt}
\end{equation}

The Hamiltonian in Eq. \eqref{eqnDiabInt} does not commute with itself at different times and is in general hard to solve. We may calcaulte the resulting unitary evolution operator with a Dyson-type series. We note that the $n$'th order term in the series scales as $\left(\dot{\gamma}T\right)^n$, which is expected to be small for small rotations, as on average $\dot{\gamma}\left(t\right)\propto\frac{\gamma\left(T\right)-\gamma\left(0\right)}{T}$. 

Thus we are content with the leading order term, and may approximate the unitary evolution operator as $U\left(T;0\right)\approx1-\frac{i}{\hbar}\int_{0}^{T}dtH_{\text{D,I}}\left(t\right)$, such that
\begin{equation}
    \varepsilon_{D}\approx\left|\int_{0}^{T}dt\dot{\gamma}e^{\frac{i}{\hbar}\int_{0}^{t}dt^{\prime}\Delta E\left(t^{\prime}\right)}\right|^{2}.\label{eqnDiabProb}
\end{equation}
Since $0\leq\gamma\leq\frac{\pi}{2}$ we can simplify Eq. \eqref{eqnDiabProb} further by using $\Delta E\left(t\right)=\frac{2\mu\delta}{\sin\left(2\gamma\right)}$ and changing the integration variable from $t$ to $\gamma$. 

We note that $\braket{\up\left(t\right)}{\partial_{t}\down\left(t\right)}=\dot{\gamma}$, so Eq. \eqref{eqnDiabProb} recovers Eq. 2.7 of \cite{Shimshoni1993}. In this context, we expect higher-order terms in the Dyson series to correspond to multiple back-and-forth tunneling events as these terms involve many of these differentiated overlaps integrated at different times.

For a linearly varying $\gamma$ we use,
\begin{equation}
    \begin{cases}
        \gamma\left(t\right)=\theta_{i}+\frac{\theta_{f}-\theta_{i}}{T}t\\
        \gamma_i=\frac{1}{2}\arctan\left(\frac{2\delta}{B_i}\right)\\
        \gamma_f=\frac{1}{2}\arctan\left(\frac{2\delta}{B_i}\right)
    \end{cases}.\label{eqnDiabLinearT}
\end{equation}
Setting this in Eq. \eqref{eqnDiabProb} we obtain,
\begin{equation}
    \varepsilon_{\text{D},\theta}\approx\frac{\theta_{f}^{2}+\theta_{i}^{2}-2\theta_{f}\theta_{i}\cos\left(\frac{\mu\delta}{\hbar}\cdot\frac{T}{\theta_{f}-\theta_{i}}\log\left(\frac{\theta_{i}}{\theta_{f}}\right)\right)}{1+\left(\frac{\mu\delta}{\hbar}\cdot\frac{T}{\theta_{f}-\theta_{i}}\right)^{2}},\label{eqnResultLinearT}
\end{equation}
which can then be constrained further to obtain the bound in Eq. \eqref{eqnRamp} of the main text. 

\begin{figure}\includegraphics[width=\columnwidth]{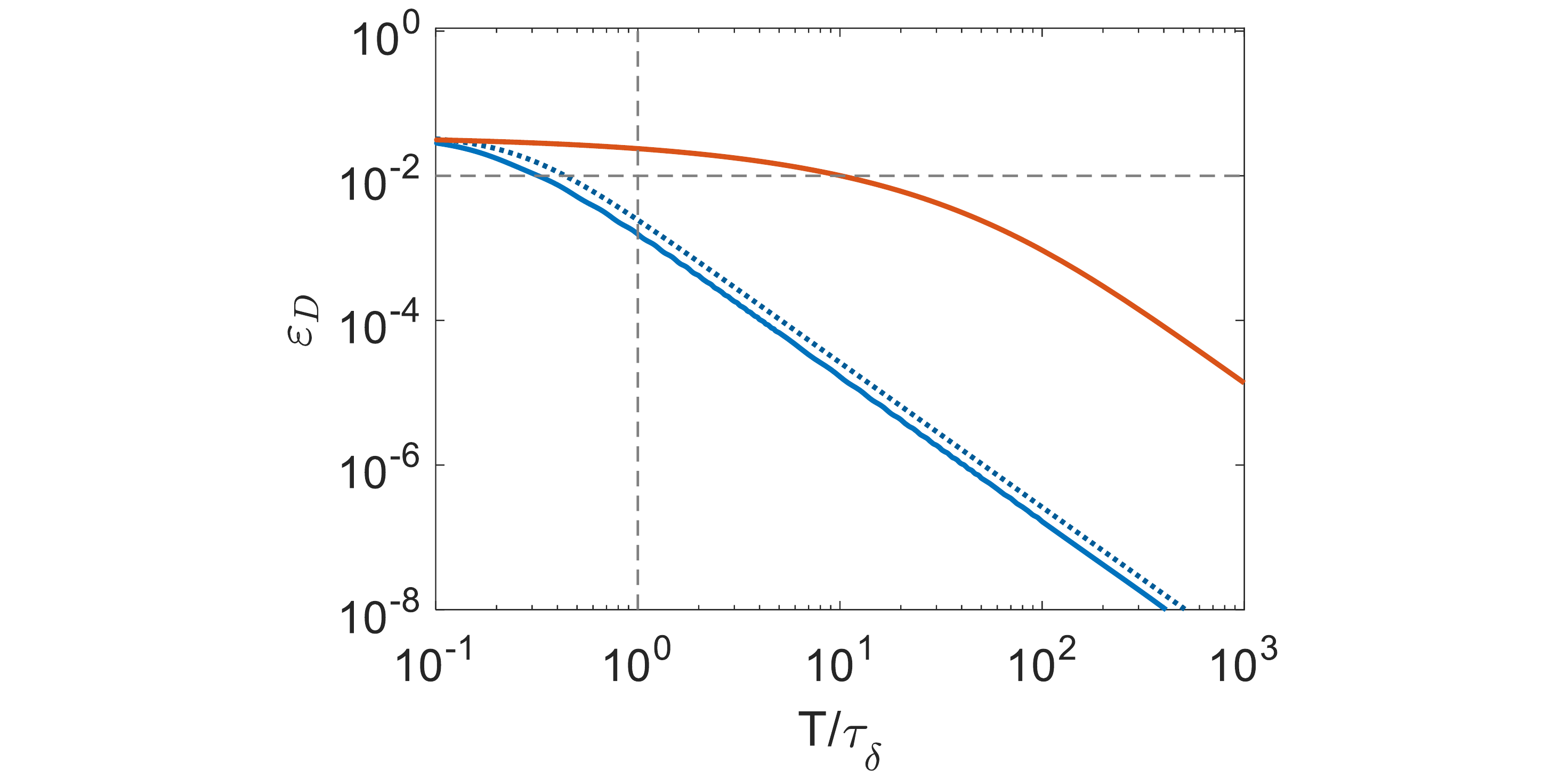}
\caption{\textbf{Probability of diabatic transition as a function of magnetic field ramp time.} The system is initialized in a clock states at $B\left(t=0\right)=B_i$, the magnetic field is then ramped down to $B\left(T\right)=B_f$ such that $\gamma\left(t\right)=\frac{1}{2}\arctan\left(\frac{2\delta}{B\left(t\right)}\right)$ varies linearly. The diabaitc transition probability is evaluated with a time-step Schrodinger's equation simulation and with the approximation in Eq. \eqref{eqnRamp}. We repeat this calculation for different field ramp times, $T$, given in units of the characteristic time $\tau_\delta=\left(\frac{\mu\delta}{\hbar}\right)^{-1}$ and with the typical parameters $B_i=100B_f=500\delta$. Clearly the approximation (dashed blue) is a good upper bound for the simulation (solid blue). The calculation is repeated for a linearly varying $B\left(t\right)$ (solid red), which is less favorable as the transition probability is higher. The dashed gray lines mark a transition probability of $\varepsilon_\text{D}=0.01$ (horizontal), and a ramp time $T=\tau_\delta$ (vertical).}\label{FigLZ}\end{figure}

To ensure our approximations are appropriate we compare our model to a simulation. Figure \ref{FigLZ} shows the diabatic transition probability due to a linear $\gamma\left(t\right)$ ramp down profile, obtained by a time-step Schrodinger's equation simulation (solid blue), compared to our approximation (dashed blue). Clearly the approximation is valid and stringent. The figure also shows the transition probability due to a linear $B\left(t\right)$ ramp down profile (solid red), which generates larger diabatic transitions and is therefore less favorable.

\begin{figure}\includegraphics[width=\columnwidth]{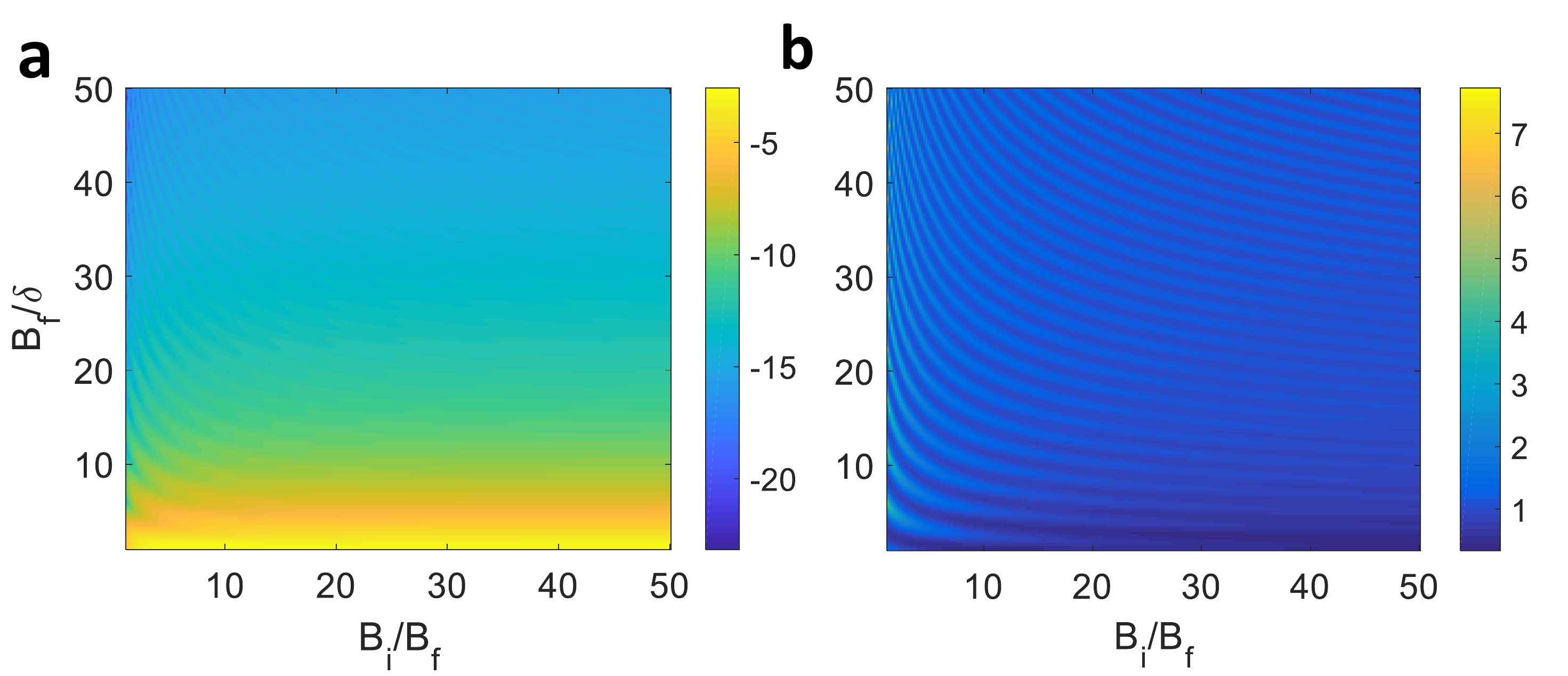}
\caption{\textbf{Simulation vs. approximation of diabatic transition probability on the parameter space  $\left(B_i/B_f,B_f/\delta\right)$ values. (a)} Diabatic transition according to Schrodinger's equation time-step simulation (log scale), showing a maximal transition probability of $0.078$. As expected, increasing both $B_i/B_f$ and $B_f/\delta$ acts to suppress this transition probability. \textbf{(b)} Approximation validity, calculated as, $\frac{\varepsilon_\text{D}-\varepsilon_\text{D,sim}}{\varepsilon_\text{D,sim}}$, according to $\varepsilon_\text{D}$ in Eq. \eqref{eqnRamp} of the main text and the simulation results in (a) (log scale). For all of the parameter space, the approximation is stringent.}\label{FigMoreLZ}\end{figure}

In Fig. \ref{FigLZ} we compare the approximation with a time-step simulation of Schrodinger's equation, for the specific choice of parameters $B_i=100 B_f=500\delta$. To ensure that the apporximation in Eq. \eqref{eqnResultLinearT} is general we set the ramp time to the $T=\tau_\delta$, and compare our approximation to a simulation for varying values of $B_i/B_f$ and $B_f/\delta$. Figure \ref{FigMoreLZ}a shows the diabatic transition probability according to the simulation, with a maximal value of $\epsilon_\text{D}=0.078$. As expected, increasing both $B_i/B_f$ and $B_f/\delta$ acts to suppress this transition. Figure \ref{FigMoreLZ}b shows the ratio approximation validity, $\frac{\varepsilon_\text{D}-\varepsilon_\text{D,sim}}{\varepsilon_\text{D,sim}}$, and indicates that the approximation is indeed stringent for the entire simulated parameter regime. 

\section{Appendix D: Numerical sensitivity optimization and self-consistency check}
\begin{figure}\includegraphics[width=\columnwidth]{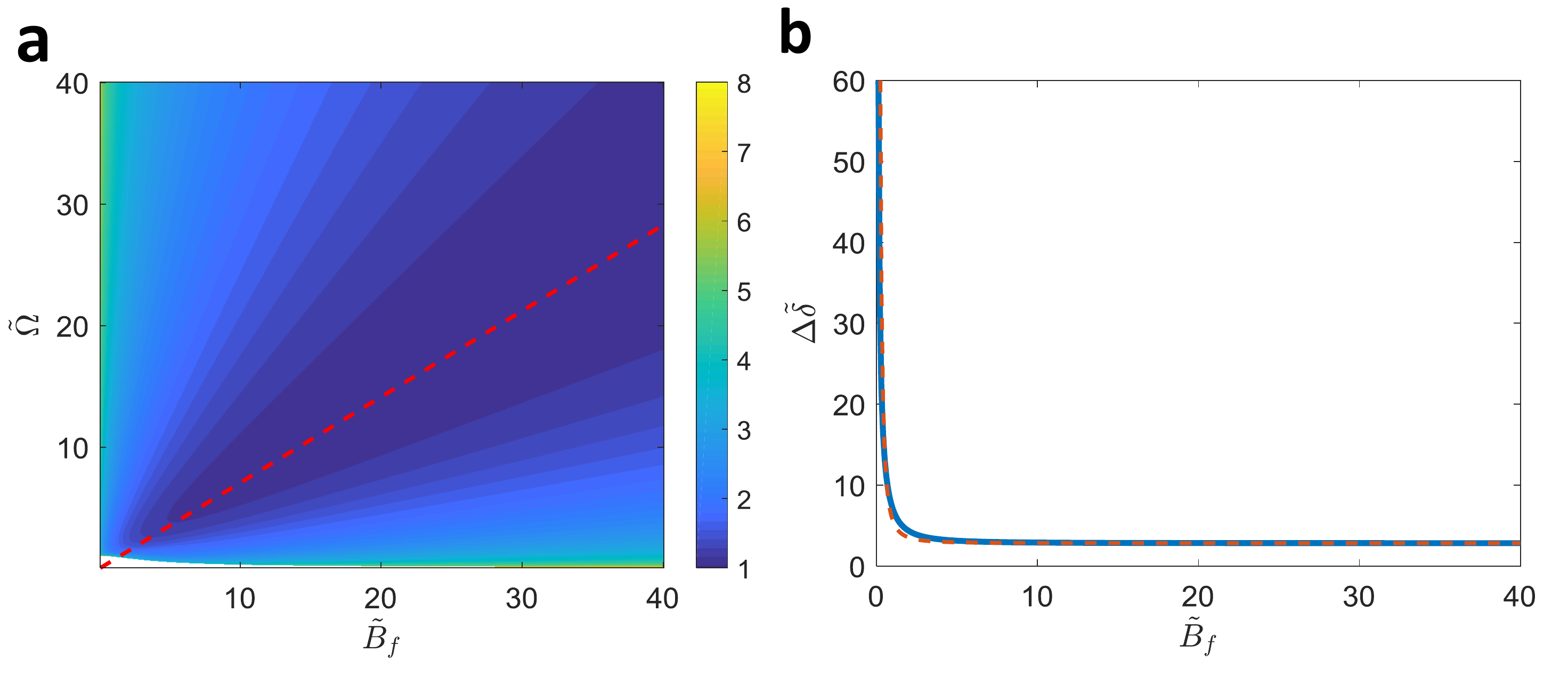}
\caption{\textbf{Numerical optimization of measurement sensitivity.} The sensitivity is numerically optimized on the  $\left(\tilde{B}_f,\tilde{\Omega}\right)$ parameter space as a function of the ramp time $\tilde{T}$. All sensitivities here are calculated for $N=1$. \textbf{(a)} Optimal sensitivity (log scale). Clearly the optimal (minimal) sensitivity lies along the $\tilde{\Omega}=\tilde{B}_f/\sqrt{2}$ line (dashed red), at $\tilde{T}\rightarrow0$, validating our derivations above. A small region in the bottom-left section (white) is excluded from the simulation since in it the errors $\epsilon_\text{D}$ and $\epsilon_{\text{P.B}}$ are no longer small and cannot be treated independently. \textbf{(b)} Optimal sensitivity as a function of $\tilde{B}_f$ (blue). Clearly the numerical optimal sensitivity fits the expression in Eq. \eqref{eqnSensSol} of the main text, shown here up to second order in $\tilde{B}_f$ (dashed red). The sensitivity quickly converges to the expected value, $\Delta\tilde{\delta}=2\sqrt{2}/\sqrt{N}$, which indicates that the "high-field" limit, used in the main text, already holds at $\tilde{B}_f\approx10$, i.e at $\delta/B_f\approx0.28$.}\label{FigNumericSens}\end{figure}

To ensure the validity of the sensitivity obtained in Eq. \eqref{eqnSensSol} of the main text, and specifically the assumption $\tilde{\Omega}\propto\tilde{B}_f$, we performed a numerical optimization of the sensitivity. We scanned the $\left(\tilde{B}_f,\tilde{\Omega}\right)$ parameter space  and numerically optimized $\tilde{T}$ in Eq. \eqref{eqnSensFull} of the main text. Figure \ref{FigNumericSens}a shows the resulting optimal sensitivity at each point of the parameter space. As expected, the optimal sensitivity lies along the $\tilde{\Omega}=\tilde{B}_f/\sqrt{2}$ line (dashed red). Figure \ref{FigNumericSens}b shows the optimal sensitivity as a function of $\tilde{B}_f$ (blue). The sensitivity converges quickly to the predicted value (dasehd red), $\Delta\tilde{\delta}=2\sqrt{2}/\sqrt{N}$.

In addition, the sensitivity obtained in Eq. \eqref{eqnSensSol} may seem inconsistent, since for $N<4$ the measurement sensitivity, $\Delta\tilde{\delta}=2\sqrt{2}/\sqrt{N}$, is larger than the measurement range $\tilde{\delta}_\text{max}=\sqrt{2}$. This occurs since we used $\delta_\text{max}$ in the expression for the diabatic error transition, $\varepsilon_\text{D}$, in Eq. \eqref{eqnRamp} of the main text. Here we repeat the sensitivity optimization analysis, in a self-consistent manner, by setting the sensitivity itself as the relevant magnetic field driving diabatic transitions, i.e $\delta\rightarrow\Delta\delta$ in Eq. \eqref{eqnRamp}. The resulting sensitivity is essentially unchanged.

We start by minimizing Eq. \eqref{eqnSensFull} of the main text with respect to $\tilde{\Omega}$. It is straight-forward to verify that the minimum is obtained at,
\begin{equation}
    \tilde{\Omega}=\sqrt{1+\frac{1}{2}\left(1-\tilde{T}\right)^{2}\tilde{B}_{f}^{2}}=\frac{1-\tilde{T}}{\sqrt{2}}\tilde{B}_{f}+\ord{\tilde{B}_{f}^{-1}}\label{eqnOmOpt}
\end{equation}
which in leading-order matches the linear scaling which is used in the main text.

By setting Eq. \eqref{eqnOmOpt} in Eq. \eqref{eqnSensFull} of the main text we obtain a quadratic equation in $\Delta\tilde{\delta}^2$, with the solution,
\begin{equation}
    \Delta\tilde{\delta}=\frac{2\sqrt{2}}{\sqrt{N}\left(1-\tilde{T}\right)}\sqrt{1+\frac{2}{\left(1-\tilde{T}\right)^2\tilde{B}_{f}^2}}+\ord{\tilde{B}_{f}^{-4}}.\label{eqnSensQuad}
\end{equation}
Here we present only the leading order contributions of this solution. The sensitivity in Eq. \eqref{eqnSensQuad} is minimized by setting $\tilde{T}=0$, yielding,
\begin{widetext}
\begin{equation}
    \Delta\tilde{\delta}=\frac{2\sqrt{2}}{\sqrt{N}}\sqrt{1+\frac{2}{\tilde{B}_{f}^{2}}+\frac{16}{N\tilde{B}_{f}^{2}}+\frac{64}{N\tilde{B}_{f}^{4}}+\frac{384}{N^{2}\tilde{B}_{f}^{4}}}
    =\frac{2\sqrt{2}}{\sqrt{N}}\left(1+\frac{1}{\tilde{B}_{f}^{2}}+\frac{8}{N\tilde{B}_{f}^{2}}\right)+\ord{\tilde{B}_{f}^{-4}},\label{eqnSensConSol}
\end{equation}
\end{widetext}
which matches the leading order solution and next order scaling of the sensitivitiy in Eq. \eqref{eqnSensSol} of the main text.

\end{document}